\documentclass{article}
\usepackage[dvips]{graphicx}
\usepackage{amssymb}
\usepackage{pstricks}
\usepackage{mathrsfs}

\newcommand{\be}{\begin{eqnarray}}
\newcommand{\ee}{\end{eqnarray}}
\newcommand{\gsim}{\mathrel{\raisebox{-.6ex}{$\stackrel{\textstyle>}{\sim}$}}}

\begin{document}

\vspace*{-0.5cm}
\begin{flushright}
OSU-HEP-08-08\\
FERMILAB-PUB-08-432-T
\end{flushright}
\vspace{0.5cm}

\begin{center}
{\Large {\bf A Light Scalar as the Messenger of Electroweak and Flavor Symmetry Breaking}}

\vspace*{1.5cm} J. D. Lykken \footnote{Email address:
lykken@fnal.gov}$^{,\dag}$, Z. Murdock
\footnote{zekemurdock@gmail.com}$^{,\ddag,}$ and  S.
Nandi\footnote{Scientific visitor at Fermilab, e-mail address:
s.nandi@okstate.edu}$^{,\dag,\ddag,}$

\vspace*{0.5cm} $^{\dag}${\it Theoretical Physics Department,
Fermilab, P.O. Box 500, Batavia, IL 60510\\}

$^{\ddag}${\it Department of Physics and Oklahoma Center for High
Energy Physics, Oklahoma State University, Stillwater, OK 74078\\}

\end{center}

\begin{abstract}

We propose a new framework for understanding the hierarchies
of fermion masses and mixings.
The masses and mixings of all Standard Model (SM)
charged fermions other than top arise from higher dimensional operators
involving a messenger scalar $S$ and flavon scalars $F_i$.
The flavons spontaneously break SM flavor symmetries at around
the TeV scale. The SM singlet scalar $S$ couples directly to the Higgs $H$ and
spontaneously breaks another $U(1)$ at the electroweak scale.
At the TeV scale, SM quarks and charged leptons have renormalizable
couplings to $S$, but not to $H$ or $F_i$. These couplings involve
new heavy vectorlike fermions. Integrating out these fermions
produces a pattern of higher dimensional operators that reproduce
the observed hierarchies of the SM masses and mixings in terms of
powers of the ``little hierarchy'': the ratio of the electroweak scale
to the flavor-breaking scale.

The framework has important phenomenological implications.
Flavor-changing neutral currents are within experimental limits but
$D^0$$-$$\bar{D}^0$ mixing and $B_s\rightarrow{\mu^+\mu^-}$ could
be close to current sensitivities.
The neutral scalar $s$ of the
messenger field mixes with the light Higgs of the SM, which can have
strong effects on Higgs decay branching fractions. The $s$ mass
eigenstate may be lighter than the Higgs, and could be detected at the
Tevatron or the LHC.

\end{abstract}

\section{Introduction}

Explaining the
fermion mass hierarchy and mixing pattern is an outstanding
challenge of particle physics \cite{Froggatt:1978nt}\cite{attempts}\cite{Nandi:2008zw}.
The fermion masses are parameterized by the Standard Model
Yukawa interactions of chiral fermions with a single Higgs doublet.
It is technically natural for the dimensionless Yukawa
couplings to take small values, since global chiral flavor symmetries
are restored (at tree level) in the limit that these couplings vanish, but it
is a total mystery why these values are spread over more than
five orders of magnitude, in a suggestive pattern of
inter-generational and intra-generational hierarchies.

Although the gauge sector of the SM is well established, little is yet
known about the Higgs sector.
Higgs physics may be much richer than the minimal SM formulation, presenting
new dynamics at the TeV scale that will be accessible to experiments
at the LHC. Most work on extended Higgs sectors has been motivated
by frameworks for understanding the naturalness and hierarchy problem
of the SM Higgs boson, but not by the hierarchy problems of the SM
flavor sector. One reason is that models that attempt to generate
the flavor-breaking patterns of the SM Yukawas from new 
TeV scale dynamics are strongly constrained by experimental searches for
flavor-changing neutral currents (FCNCs) and charged lepton
flavor violation (CLFV).

The top quark Yukawa coupling has a value close to one,
suggesting that a SM Yukawa coupling is the correct explanation
for the top mass. The smallness of the other Yukawas suggests
that some or all of the other quarks and the charged leptons do not couple directly
to the electroweak symmetry breaking order parameter, which in the SM
is represented by the vacuum expectation value (vev) of the
Higgs scalar. Thus a good starting
point to construct theories of flavor is to specify a field or mechanism
to act as the messenger of electroweak symmetry breaking to the
other quarks and leptons.

One simple choice for a messenger is a TeV mass scalar
leptoquark, postulated to have a renomalizable coupling between
the top quark and the SM leptons \cite{Dobrescu:2008sz,Balakrishna:1987qd}.
Radiative corrections can then
generate a natural hierarchy of fermion masses related to powers
of a loop factor.

An even simpler choice for a messenger is an electroweak mass scalar
that transforms as a SM singlet and extends the Higgs sector of the SM.
In this work, we explore this idea of an extended Higgs sector
related to the generation of the fermion mass hierarchy. We
present a simple framework where the Higgs doublet $H$ couples directly to
a complex scalar $S$ that is a SM singlet and is charged under
a new local $U(1)_S$ symmetry carried by a vector boson $Z'$.
All of the SM fermions are singlets under
this new $U(1)_S$ (apart from small effects from $Z-Z'$ mixing),
which is broken spontaneously at the electroweak scale by
the vacuum expectation value of $S$.

In our framework the singlet scalar $S$ is the messenger to SM fermions of both flavor
breaking and electroweak symmetry breaking. All SM fermions apart from
the third generation quark doublet $q_{3L}$ and right-handed top $u_{3R}$
are assumed to carry a nonzero charge under a gauged chiral flavor symmetry
forbidding all SM dimension 4 Yukawa couplings except that of the
top quark. We assume that the flavor symmetry is spontaneously broken at a
scale $\gsim 1$ TeV by the vacuum expectation of one or more complex scalar ``flavon''
fields $F_i$. The flavor charges of the SM fermions forbid any
dimension 4 couplings to either $F_i$ or to the Higgs field $H$.

We introduce new fermions
that are vectorlike under both the SM gauge symmetries and $U(1)_S$;
these fermions naturally acquire masses $\gsim$ TeV that we will
generically denote as $M$, and have
dimension 4 couplings to both $F_i$ and to $H$. Integrating out
these heavy fermions gives higher dimension effective couplings
of the SM fermions to $H$ that replace the role of Yukawa couplings in the SM.
These couplings contain explicit flavor breaking in the form of
$\langle F_i \rangle /M$, which we take to be of order 1, as well as being
suppressed by powers of $S^\dagger S / M^2$, whose vev we take to be
of order 1/50.

In our framework all of the observed SM fermion mass hierarchies are
generated from powers of $\langle S \rangle /M \sim 1/7$, which is
essentially the ratio of the electroweak scale to the TeV scale,
often called the ``little hierarchy''. We can be agnostic about
the source of the little hierarchy itself, since many possibilities
have been proposed. The additional challenge of our framework
is to achieve simultaneously the appropriate flavon physics at the TeV scale.

Models in our framework have, in addition to the SM particle content,
a light singlet scalar $s$ that mixes with the Higgs boson $h$.
Exchanges of $s$ between SM fermions are a new source of FCNC.
There is an extra $Z'$ at the electroweak (EW) scale, but apart from small
$Z-Z'$ mixing effects it does not couple to SM fermions.
There may be other $Z'$s and one or more flavon scalars at the TeV scale.
We predict a host of new heavy fermions around
the TeV scale; these are also a source of new FCNC and CLFV effects.
We show that flavon charge patterns that reproduce the observed
SM fermion masses and mixings also supply enough extra suppression
of FCNC and CLFV effects to satisfy current experimental bounds.

In addition to explaining the hierarchy of fermion masses
and mixings, models in our framework have many interesting phenomenological implications.
Mixing of the singlet $s$ with the Higgs boson $h$ can cause
large deviations from the SM predictions for the Higgs decay branching
fractions, potentially observable at the Tevatron or LHC.
The $s$ mass eigenstate itself will also
be produced at the LHC, and could be confused with $h$
if it turns out to be the lightest mass eigenstate. While new
FCNC effects are suppressed, we predict contributions to
$D^0$$-$$\bar{D}^0$ mixing, $B_s\rightarrow{\mu^+ \mu^-}$,
that are close to the current value or limit. The exotic top
quark decays $t\rightarrow{ch}$ and $t\rightarrow{cs}$ can
have branching fractions on the order of $10^{-3}$.

Our paper is organized as follows. In section 2, we present the
basic outline of our framework. In section 3, we discuss the constraints
on the model parameters from the low energy phenomenology. Section 4
contains the phenomenological implications and predictions of the
model, especially for the new top decays and Higgs signals at the
Tevatron and LHC. In section 5, we outline a possible ultraviolet
completion realizing our proposal. Section 6 contains our
conclusions and further discussion.

\section{Model and formalism}

We extend the gauge symmetry of the SM by a $U(1)_S$
local symmetry and an additional local flavon symmetry which in the
simplest case would be a $U(1)_F$.
All of the SM fermions are neutral with respect to
$U(1)_S$, while all of the SM fermions apart from
the third generation quark doublet $q_{3L}$ and right-handed top $u_{3R}$
are charged under the chiral $U(1)_F$. We introduce a complex scalar field $S$
which has charge 1 under $U(1)_S$, is neutral under the flavon symmetry, and is
a SM singlet. We also introduce one or more complex scalar fields $F_i$, the ``flavon'' scalars.
In the simplest case there would be a single flavon scalar $F$
that has charge 1 under $U(1)_F$, is neutral under $U(1)_S$, and is
a SM singlet. The Higgs field $H$ is taken as neutral under $U(1)_S \times U(1)_F$.
We assume that the flavon charges of the SM fermions
are such that only the top quark has an allowed dimension 4 Yukawa
interaction.

The $S$ field is assumed to develop a vev that spontaneously breaks
the $U(1)_S$ symmetry. In frameworks where the little hierarchy
between the electroweak scale and the TeV scale is generated, this could
occur naturally by extending the Higgs sector to include $S$, with a
mixed potential. The pseudoscalar component of $S$ is then ``eaten''
to give mass to the $U(1)_S$ $Z'$ gauge boson. Notice that the vev
of $S$ does not in itself break any of the global flavor symmetries of
the Yukawa-less SM; $S$ is only a messenger of flavor breaking, just as
it is also a messenger of electroweak breaking. This is the fundamental
distinction that allows $S$ to exist at the electroweak scale without
inducing unacceptably large flavor violating effects.

The
flavon scalars $F_i$ are assumed to develop vevs that spontaneously break
the local flavon symmetry at the TeV scale, with the pseudoscalar components of the $F_i$
eaten to give the flavon gauge bosons mass. To preserve the little
hierarchy, we assume that the direct mixing between the $F_i$ and the
extended Higgs sector is negligible.

In this framework the Yukawa interactions of the lighter quarks and leptons
are replaced by higher dimension operators that couple these fermions
to $H$, $S$, and the $F_i$. As we will show later in an explicit example, these
can be generated as effective couplings by integrating out new heavy fermions
at the TeV scale. These effective couplings should respect all of the SM gauge
symmetries, as well as $U(1)_S$ and the flavon symmetries. In particular, the $U(1)_S$
charged field $S$ can only appear as powers of $S^\dagger S / M^2$, where
$M$ denotes a generic TeV scale parameter. Powers of $F_i/M$ and $F_i^\dagger /M$
can also appear, but the exact form depends on the flavon charge assignments
of the SM fermions. Since we will assume that vevs of the $F_i$ are of order $M$,
we can absorb the $F_i/M$ dependence into the dimensionless
complex couplings $h_{ij}$, where $i$, $j$ are generation labels;
all these couplings we will then take to be of order 1.

The observed SM fermion mass hierachy is generated
from the following low energy effective interactions:

\begin{eqnarray}
{\cal L}^{\rm Yuk} &=& h_{33}^u \overline{q}_{3L} u_{3R} \bar{H} +
\left({S^\dagger S \over M^2}\right) \left(h_{33}^d
\overline{q}_{3L} d_{3R} H + h_{22}^u \overline{q}_{2L} u_{2R}
\bar{H}+h_{23}^u \overline{q}_{2L} u_{3R} \bar{H}
+ h_{32}^u \overline{q}_{3L} u_{2R} \bar{H}\right) \nonumber \\
&&+ \left({S^\dagger S \over M^2}\right)^2
\left(h_{22}^d\overline{q}_{2L} d_{2R} H + h_{23}^d
\overline{q}_{2L} d_{3R} H + h_{32}^d \overline{q}_{3L} d_{2R} H +
h_{12}^u\overline{q}_{1L}u_{2R}\bar{H}
+ h_{21}^u\overline{q}_{2L}u_{1R} \bar{H} \right.  \nonumber \\
&&+ \left.  h_{13}^u\overline{q}_{1L} u_{3R} \bar{H} + h_{31}^u
\overline{q}_{3L} u_{1R} \bar{H} \right) + \left({S^\dagger S
\over M^2}\right)^3 \left(h_{11}^u \overline{q}_{1L} u_{1R}
\bar{H}
+ h_{11}^d\overline{q}_{1L} d_{1R} H  \right.  \nonumber \\
&&+ \left. h_{12}^d\overline{q}_{1L} d_{2R} H + h_{21}^d
\overline{q}_{2L} d_{1R} H + h_{13}^d \overline{q}_{1L} d_{3R} H +
h_{31}^d \overline{q}_{3L} d_{1R} H \right ) + h.c. \label{ONE}
\end{eqnarray}

Note that the above interactions are very similar to those proposed
in reference \cite{bn}, except our interactions involve suppression
by powers of $\left({S^\dagger S \over M^2}\right)$, instead of
$\left({H^\dagger H \over M^2}\right)$. We will refer to this as the
Babu-Nandi texture. The hierarchies among the fermion masses and
mixings are obtained from a single small dimensionless parameter,

\begin{equation}
\epsilon \equiv { v_s \over M},
\label{TWO}
\end{equation}
where $v_s$ is the vev of $S$.
As was shown in \cite{bn}, a good fit to the observed fermion masses and
mixings is obtained with $\epsilon \sim 0.15$. The couplings
$h_{ij}$ are all of order one; the largest coupling needed is $h^u_{23}
=1.4$, while the smallest coupling needed is $h^u_{22}=0.14$.

The Babu-Nandi texture is not unique, and it does not predict any precise
fermion mass relations,
since there are slightly more unspecified order 1 parameters than there are
Yukawa parameters in the SM.

\subsection{Fermion masses and CKM mixing}

The gauge symmetry of our model is the usual $SU(3)_c\times SU(2)_L \times U(1)_Y$ of the SM, 
plus two
additional local symmetries: $U(1)_S$ and the flavon symmetry. 
The SM symmetry is broken spontaneously
by the usual Higgs doublet $H$ at the electroweak scale. We assume that the
extra $U(1)_S$ symmetry is also broken spontaneously at the electroweak scale
by a SM singlet complex scalar field $S$. The flavon symmetry, $U(1)_F$ in the simplest case,
is broken spontaneously above a TeV
by a SM singlet scalar flavon field $F$. The pseudoscalar part of
the complex scalar field $S$ is absorbed by the $Z'$ gauge boson $U(1)_S$ to get its mass.
Thus after symmetry breaking the remaining scalars at the electroweak scale are
neutral bosons $h$ and
$s$. Parameterizing the Higgs doublet and singlet in the unitary
gauge as

\begin{equation}
H = \left(\matrix{0 \cr \frac{h}{\sqrt{2}}+v}\right)~~S = \left(\frac{s}{\sqrt{2}}+v_s\right),
\label{THREE}
\end{equation}
with $v \simeq 174$ GeV, and defining an additional small parameter

\begin{equation}
\beta \equiv { v \over M},
\label{FOUR}
\end{equation}
we obtain, from eqs. (\ref{ONE}-\ref{FOUR}) the following mass matrices for the up and down quark sector:

\begin{eqnarray}
M_u = \left(\matrix{h_{11}^u \epsilon^6 & h_{12}^u \epsilon^4 &
h_{13}^u \epsilon^4 \cr h_{21}^u\epsilon^4 & h_{22}^u \epsilon^2 &
h_{23}^u \epsilon^2 \cr h_{31}^u \epsilon^4 & h_{32}^u \epsilon^2 &
h_{33}^u}\right)v, ~~~~~ M_d = \left(\matrix{h_{11}^d \epsilon^6 &
h_{12}^d \epsilon^6 & h_{13}^d \epsilon^6 \cr h_{21}^d\epsilon^6 &
h_{22}^d \epsilon^4 & h_{23}^d \epsilon^4 \cr h_{31}^d \epsilon^6 &
h_{32}^d \epsilon^4 & h_{33}^d \epsilon^2}\right)v~.
\label{FIVE}
\end{eqnarray}
The charged lepton mass matrix is obtained from $M_d$ by replacing
the couplings $h_{ij}$ appropriately. Note that these mass matrices
are the same as in \cite{bn}, and as was shown there, good fits to the
quark and charged lepton masses, as well as the CKM mixing angles are
obtained by choosing $\epsilon\sim 0.15$, and all the couplings
$h_{ij}$ of order one. To leading order in $\epsilon$, the
fermion masses are given by
\be\label{eqn:diagmasses}
(m_t,\;m_c\;,m_u) &\simeq& (\vert h_{33}^u\vert,\; \vert h_{22}^u\vert\epsilon^2,\;
\vert h_{11}^u - h_{12}^uh_{21}^u/h_{22}^u\vert \epsilon^6)\,v\;,\nonumber\\
(m_b,\; m_s,\; m_d) &\simeq& (\vert h_{33}^d\vert\epsilon^2,\; \vert
h_{22}^d\vert\epsilon^4,\;
\vert h_{11}^d\vert\epsilon^6)\,v\;,\\
(m_{\tau},\;m_{\mu},\;m_e) &\simeq& (\vert h_{33}^\ell\vert\epsilon^2,\;
\vert h_{22}^\ell\vert\epsilon^4,\; \vert h_{11}^\ell\vert\epsilon^6)\,v \; ,\nonumber
\ee
while the quark mixing angles are
\be\label{eqn:mixings}
\vert V_{us}\vert &\simeq& \left\vert \frac{h_{12}^d}{h_{22}^d} - \frac{h_{12}^u}{h_{22}^u} \right\vert\epsilon^2 \; ,\nonumber\\
\vert V_{cb}\vert &\simeq& \left\vert  \frac{h_{23}^d}{h_{33}^d} - \frac{h_{23}^u}{h_{33}^u} \right\vert\epsilon^2 \; , \\
\vert V_{ub}\vert &\simeq& \left\vert \frac{h_{13}^d}{h_{33}^d} - \frac{h_{12}^uh_{23}^d}{h_{22}^uh_{33}^d}
- \frac{h_{13}^u}{h_{33}^u} \right\vert\epsilon^4 \; . \nonumber
\ee

Generically all of the $h_{ij}$ can be nonvanishing, but in a particular
ultraviolet (UV) completion flavon charge conservation may push some of
them to higher order in $\epsilon$ or to vanish altogether. However
from (\ref{eqn:diagmasses}) and (\ref{eqn:mixings}) we see that the
Babu-Nandi texture is rather robust:
the only flavor off-diagonal couplings needed to reproduce the
observed mixings are one or more of $h_{12}^d$, $h_{12}^u$, one or more of
$h_{23}^d$, $h_{23}^u$, and one or more of $h_{13}^d$, $h_{13}^u$;
the rest can either vanish or appear at higher order in $\epsilon$.

\subsection{Yukawa interactions and FCNC}

Our model has flavor changing neutral current interactions in the
Yukawa sector. Using eqs.(1-4), the Yukawa interaction matrices
$Y^{h}_u$, $Y^{h}_d$, $Y^{s}_u$, $Y^{s}_d $ for the up and down
sector, for $h^0$ and $s^0$ fields are obtained to be

\begin{eqnarray}
\sqrt{2} Y^{h}_u = \left(\matrix{h_{11}^u \epsilon^6 & h_{12}^u
\epsilon^4 & h_{13}^u \epsilon^4 \cr h_{21}^u\epsilon^4 & h_{22}^u
\epsilon^2 & h_{23}^u \epsilon^2 \cr h_{31}^u \epsilon^4 & h_{32}^u
\epsilon^2 & h_{33}^u}\right), ~~~~~ \sqrt{2} Y^{h}_d =
\left(\matrix{h_{11}^d \epsilon^6 & h_{12}^d \epsilon^6 & h_{13}^d
\epsilon^6 \cr h_{21}^d\epsilon^6 & h_{22}^d \epsilon^4 & h_{23}^d
\epsilon^4 \cr h_{31}^d \epsilon^6 & h_{32}^d \epsilon^4 & h_{33}^d
\epsilon^2}\right), \label{SIX}
\end{eqnarray}
with the charged lepton Yukawa coupling matrix $Y_\ell$ obtained
from $Y_d$ by replaing $h_{ij}^d \rightarrow h_{ij}^\ell$.

\begin{eqnarray}
\sqrt{2} Y^{s}_u = \left(\matrix{6h_{11}^u \epsilon^5\beta &
4h_{12}^u \epsilon^3\beta & 4h_{13}^u \epsilon^3\beta \cr
4h_{21}^u\epsilon^3\beta & 2h_{22}^u \epsilon\beta & 2h_{23}^u
\epsilon\beta \cr 4h_{31}^u \epsilon^3\beta & 2h_{32}^u
\epsilon\beta & 0}\right), ~~~~~ \sqrt{2} Y^{s}_d =
\left(\matrix{6h_{11}^d \epsilon^5\beta & 6h_{12}^d \epsilon^5\beta
& 6h_{13}^d \epsilon^5\beta \cr 6h_{21}^d\epsilon^5\beta & 4h_{22}^d
\epsilon^3\beta & 4h_{23}^d \epsilon^3\beta \cr 6h_{31}^d
\epsilon^5\beta & 4h_{32}^d \epsilon^3\beta & 2h_{33}^d
\epsilon\beta}\right), \label{SEVEN}
\end{eqnarray}
with the charged lepton Yukawa coupling matrix $Y_\ell$ obtained
from $Y_d$ by replaing $h_{ij}^d \rightarrow h_{ij}^\ell$.

There are several important features that distinguish our model from
the proposals in \cite{bn,gl,Dorsner:2002wi}:

i) Note, from eqs.(\ref{FIVE}) and (\ref{SIX}), in our model, the
Yukawa couplings of h to the SM fermions are exactly the same as in
the SM. This is because the fermion mass hierarchy in our model is
arising from $\left({S^\dagger S \over M^2}\right)$. This is a
distinguishing feature of our model from that proposed in
\cite{bn,gl} where the Yukawa couplings of $h$ are flavor dependent,
because the hierarchy there arises from $\left({H^\dagger H \over
M^2}\right)$.

ii) In our model, we have an additional singlet Higgs boson whose
couplings to the SM fermions are flavor dependent as given in eq.
(\ref{SEVEN}). Again, this is because the hierarchy in our model
arises from $\left({S^\dagger S \over M^2}\right)$. In particular,
$s^0$ does not couple to the top quark, and its dominant fermionic
coupling is to the bottom quark. This will have interesting
phenomenological implications for the Higgs searches at the LHC.

iii) We note from eqs. (\ref{FIVE}-\ref{SIX}) that the mass matrices
and the corresponding Yukawa coupling matrices for $h$ are proportional
as in the SM. Thus there are no flavor changing Yukawa interactions
mediated by $h$. However, this is not true for the Yukawa interactions
of the singlet Higgs as can be seen from eqs. (\ref{FIVE}) and
(\ref{SEVEN}). Thus $s$ exchange will lead to flavor violation in the
neutral Higgs interactions.

\subsection{Higgs sector and the $Z'$}

The Higgs potential of our model, consistent with the SM and the
extra $U(1)_S$ symmetry, can be written as

\begin{eqnarray}
 V(H,S) = -\mu^{2}_H (H^{\dag} H) - \mu^{2}_S (S^{\dag} S)
 + \lambda_H (H^{\dag}H)^2 + \lambda_S (S^{\dag} S)^2
 +\lambda_{HS}(H^{\dag}H)(S^{\dag} S).
\label{EIGHT}
\end{eqnarray}

Note that after absorbing the three components of H in $W^{\pm}$ and
Z, and the pseudoscalar component of S in $Z'$, we are left with
only two scalar Higgs, $h^0$ and $s^0$. The squared mass matrix in
the $(h^0, s^0)$ basis is given by
\begin{equation}
{\cal M}^2 = 2 v^2\left(\matrix{
    2\lambda_H          &  \lambda_{HS} \alpha \cr
     \lambda_{HS} \alpha    & 2\lambda_S \alpha^2  \cr
        }\right),
\label{NINE}
\end{equation}
where $\alpha=v_s/v$.

The mass eigenstates $h$ and $s$ can be written as

\begin{eqnarray}
 h^0 &=& h \cos\theta + s \sin\theta, \nonumber \\
 s^0 &=& - h \sin\theta + s \cos\theta,
\label{TEN}
\end{eqnarray}
 where $\theta$ is the  mixing angle in the Higgs sector.

 In the Yukawa interactions discussed above, as well as in the gauge
 interactions involving the Higgs fields, the fields appearing are
 $h^0$ and $s^0$, and these can be expressed in terms of $h$ and $s$
 using eq. (\ref{TEN}).

 The mass of the $Z'$ gauge boson is given by

 \begin{equation}
 m^{2}_{Z'} = 2 g^{2}_E  v^{2}_s
 \label{ELEVEN}
 \end{equation}

Note that the $Z'$ does not couple to any SM particles directly. 
The $Z'$ coupling to the SM particles will be only via dimension
six or higher operators. Such couplings will be generated by the
vectorlike fermions in the model to be discussed in section 5.

\section {Phenomenological Implications: Constraints from existing data}

In this section, we discuss the constraints on our model from the
existing experimental results. As can be seen from eq. (\ref{SEVEN}), the
exchange of $s$ gives rise to tree level FCNC processes. This will
cause $K^0$$-$$\bar{K}^0$ mass splitting, $D^0$$-$$\bar{D}^0$ mixing, $K_L
\rightarrow \mu^+\mu^-$, $B^{0}_{s} \rightarrow \mu^+\mu^-$, as well
as contributions to the electric dipole moment (EDM) of neutron
and electron, and other rare processes that we discuss below.

\subsection{$\mathbf{K^0-\bar{K}^0}$ mixing} In our model, this arises from
the tree level $s$ exchange between $d\bar{s}$ and $\bar{s}d$, and is
proportional to $\beta^2 \epsilon^{10}$. Taking $\beta\sim \epsilon$
$\sim 0.15$, and the values of the couplings $h^{d}_{12}$ and
$h^{d}_{21}$ to be of order 1, the contribution to $\Delta m_K^{\rm
Higgs} \simeq 10^{-16}$ to $10^{-17}$ GeV, for an $s$ mass of 100 GeV. The
experimental value of $\Delta m_K$ is $3.5 \times 10^{-15}$ GeV
\cite{pdg}. Thus, since the contribution goes like $m_s^{-4}$, s can
be lighter than 100 GeV. Note that $\epsilon = v_s/M$ is fixed
to be $\sim 0.15$ to explain fermion mass hierarchy and the CKM
mixing. However, $\beta = v/M$ is a parameter in our model. Although
the $\Delta m_K$ constraint allows a somewhat larger value of $\beta$, we
shall see that $D^0-\bar{D^0}$ mixing constrains $\beta \sim
\epsilon$.

\subsection{$\mathbf{D^0-\bar{D}^0}$ mixing}\label{sec:3.2}
This contribution is again due to the
tree level $s$ exchange between $u\bar{c}$ and $\bar{u}c$, and is
proportional to $\beta^2\epsilon^6$, and hence is enhanced compared
to $\Delta m_K$. Again, taking the couplings $h^{u}_{12}$ and
$h^u_{21}$ to be of order one and $\beta\sim \epsilon$, we get $\Delta
m_D \sim 10^{-14}$ GeV for $m_s = 100$ GeV. This is to be compared
with the current experimental value of $1.6\times 10^{-14}$ GeV
\cite{pdg,cdf}. Thus $\Delta m_D$ gives a much stronger restriction on
the model parameters. $\beta$ can not be much larger than
$\epsilon$, and $s$ can not be much lighter than 100 GeV. If our
proposal is correct, an electroweak singlet scalar should be observed at
the LHC.

\subsection{Other rare processes} In our model, tree level $s$
exchange between $d\bar{s}$ and $\mu^+ \mu^-$ will contribute to
$K_L \rightarrow \mu^+\mu^-$. This contribution is proportional to
$\beta^2\epsilon^{10}$, and leads to a contribution to this branching
ratio $\sim 10^{-14}$ for $\beta \sim \epsilon$ and $m_s\sim 100$
GeV. This is very small compared to the current experimental value of $\sim 6.9 \times 10^{-9}$
\cite{pdg}. Similarly, the contribution to the other rare processes
such as $K_L \rightarrow \mu e$, $K \rightarrow \pi \bar{\nu} \nu$,
$\mu \rightarrow e \gamma$, $\mu \rightarrow 3e$, $B_d-\bar{B_d}$
mixing, etc are several orders of magnitude below the corresponding
experimental limits.

\subsection{Constraint on the mass of $s$} Experiments at LEP2 have
set a lower limit of $114.4$ GeV for the mass of the SM Higgs boson,
from nonobservation of the
associated production $e^+ e^- \rightarrow Zh$. In our model, since
the singlet Higgs can mix with the doublet $h$, there will be a limit
for $m_s$ depending on the value of the mixing angle, $\theta$. For
$\sin^2\theta\ge 0.25$, the bound of $114.4$ applies also for $m_s$
\cite{lpew}. However, $s$ can be lighter if the mixing is small.

\subsection{Constraint on the mass of the $Z'$} We have assumed that
the extra $U(1)$ symmetry in our model is spontaneously broken at
the EW scale. But the corresponding gauge coupling, $g_E$ is
arbitrary and hence the mass of $Z'$ is not determined in our model.
However, very accurately measured $Z$ properties at LEP1 put a
constraint on the $Z-Z'$ mixing to be $\sim 10^{-3}$ or smaller
\cite{pdg,Langacker:2008yv}. In our model, the $Z'$ does not couple to any SM
particle directly. $Z-Z'$ mixing can take place at the one loop
level with the new vectorlike fermions in the loop. The mixing angle
is

\begin{equation}
    \theta_{ZZ'}\sim \frac{g_Z g_E}{16 \pi^2} \left(\frac{m_Z}{M}\right)^2,
\label{TWELVE}
\end{equation}
where M is the mass of the vectorlike fermions with masses in the
TeV scale. Even with $g_E \sim 1$, we get $\theta_{ZZ'} \sim 10^{-4}$ or
less. Thus there is no significant bound for the mass of this $Z'$
from the LEP1. This $Z'$ can couple to the SM particles via
dimension six operators with the interaction of the form

\begin{equation}
 L = \frac{\bar{\psi}_L\sigma^{\mu\nu} \psi_R H  Z'^{\mu\nu}}{M^2}
\; .
\label{THIRTEEN}
\end{equation}
As was shown in \cite{dob}, no significant bound on $m_{Z'}$
 emerges from these interactions.

 \section{Phenomenological Implications: New physics
 signals}

 Motivated to explain the observed mass hierarchy in the fermion
 sector, we have constructed a model which has a complex singlet Higgs
 (in addition to the usual doublet), a new $U(1)_S$ gauge symmetry
 at the EW scale, and a new set of vectorlike fermions at the TeV
 scale. Thus our model has new particles such as a scalar Higgs and
 a new $Z'$ boson at the EW scale, and heavy vectorlike quarks and
 leptons. The model has many phenomenological
 implications for the production and decays of the Higgs bosons,
 top quark physics, a new scenario for $Z'$ physics, and the
 production and decays of the vectorlike fermions.

 \subsection{Higgs signals}

 \subsubsection{Higgs coupling to the SM fermions} As can be seen from (\ref{SIX}), 
the couplings of the doublet Higgs $h$ to the SM fermions
 are identical to that in the SM, whereas the couplings of the
 singlet Higgs have a different flavor dependence. In particular, the singlet
 Higgs $s$ does not couple to the top quark, whereas its couplings to
 $(b,\tau; c,s,\mu; u,d,e)$ involve the flavor dependent factors
 $(2,2; 2,4,4; 6,6,6)$ respectively, in the limit
 of zero mixing between $h$ and $s$. Including the mixing, these factors
 will be modified. Thus our model
 will be distinguished from the SM by the fact that the Higgs has
nonstandard couplings
to fermions predicted in terms of two model parameters: the
ratio of vevs $\alpha$ and the mixing angle $\theta$.

 \subsubsection{Higgs decays} The couplings of the Higgs bosons $h$ and
 $s$ to the fermions and the gauge bosons can be obtained from
 eqns. (\ref{SIX}) and (\ref{SEVEN}), and are given in Table 1.
\begin{table}
\begin{center}
    \begin{tabular}{|l|c||l|c|}
     \hline
        \bf{Interaction} & \bf{Coupling} & \bf{Interaction} & \bf{Coupling}\\
$s\rightarrow u\overline{u}$ & $\frac{m_u}{v\sqrt{2}}\left(\sin\theta+\frac{6\cos\theta}{\alpha}\right)$ & $h\rightarrow u\overline{u}$ & $\frac{m_u}{v\sqrt{2}}\left(\cos\theta-\frac{6\sin\theta}{\alpha}\right)$\\
$s\rightarrow d\overline{d}$ & $\frac{m_d}{v\sqrt{2}}\left(\sin\theta+\frac{6\cos\theta}{\alpha}\right)$ & $h\rightarrow d\overline{d}$ & $\frac{m_d}{v\sqrt{2}}\left(\cos\theta-\frac{6\sin\theta}{\alpha}\right)$\\
$s\rightarrow \mu^+\mu^-$ & $\frac{m_\mu}{v\sqrt{2}}\left(\sin\theta+\frac{4\cos\theta}{\alpha}\right)$ & $h\rightarrow \mu^+\mu^-$ & $\frac{m_\mu}{v\sqrt{2}}\left(\cos\theta-\frac{4\sin\theta}{\alpha}\right)$\\
$s\rightarrow s\overline{s}$ & $\frac{m_s}{v\sqrt{2}}\left(\sin\theta+\frac{4\cos\theta}{\alpha}\right)$ & $h\rightarrow s\overline{s}$ & $\frac{m_s}{v\sqrt{2}}\left(\cos\theta-\frac{4\sin\theta}{\alpha}\right)$\\
$s\rightarrow \tau^+\tau^-$ & $\frac{m_\tau}{v\sqrt{2}}\left(\sin\theta+\frac{2\cos\theta}{\alpha}\right)$ & $h\rightarrow \tau^+\tau^-$ & $\frac{m_\tau}{v\sqrt{2}}\left(\cos\theta-\frac{2\sin\theta}{\alpha}\right)$\\
$s\rightarrow c\overline{c}$ & $\frac{m_c}{v\sqrt{2}}\left(\sin\theta+\frac{2\cos\theta}{\alpha}\right)$ & $h\rightarrow c\overline{c}$ & $\frac{m_c}{v\sqrt{2}}\left(\cos\theta-\frac{2\sin\theta}{\alpha}\right)$\\
$s\rightarrow b\overline{b}$ & $\frac{m_b}{v\sqrt{2}}\left(\sin\theta+\frac{2\cos\theta}{\alpha}\right)$ & $h\rightarrow b\overline{b}$ & $\frac{m_b}{v\sqrt{2}}\left(\cos\theta-\frac{2\sin\theta}{\alpha}\right)$\\
$s\rightarrow t\overline{t}$ & $\frac{m_t}{v\sqrt{2}}\sin\theta$ & $h\rightarrow t\overline{t}$ & $\frac{m_t}{v\sqrt{2}}\cos\theta$\\
$s\rightarrow ZZ$ & $\frac{2 m_Z^2}{v\sqrt{2}}\sin\theta$ & $h\rightarrow ZZ$ & $\frac{2 m_Z^2}{v\sqrt{2}}\cos\theta$\\
$s\rightarrow Z'Z'$ & $\frac{m_{Z'}^2}{v\alpha\sqrt{2}}\cos\theta$ & $h\rightarrow Z'Z'$ & $\frac{m_{Z'}^2}{v\alpha\sqrt{2}}\sin\theta$\\
$s\rightarrow W^+W^-$ & $\frac{2 m_W^2}{v\sqrt{2}}\sin\theta$ & $h\rightarrow W^+W^-$ & $\frac{2 m_W^2}{v\sqrt{2}}\cos\theta$\\
 & & $h\rightarrow ss$ & $\lambda_{\mathrm{hss}}$\\
    \hline
    \end{tabular}
    \caption{Yukawa and gauge couplings of $h$ and $s$.}
\end{center}
\label{table1}
\end{table}

The coupling of $h$ to $s$ given by:
\begin{eqnarray*}
\lambda_{\mathrm{hss}} &=&\frac{m_h^2}{4 v}\left\{(1-\mu)\sin 2\theta\left[\cos^3\theta-\alpha\sin^3\theta+\sin 2\theta(\alpha\cos\theta-\sin\theta)\right]+\right.\\
&&\left. 3\sin 2\theta\left[\sin\theta\left(1+\mu-(1-\mu)\cos 2\theta\right)-\cos\theta\left(1+\mu-(1-\mu)\cos 2\theta\right)/\alpha\right]\right\}
\; ,
\end{eqnarray*}
where $\mu=m_s^2/m_h^2$.

\begin{figure}
    \begin{center}
        \includegraphics[scale=1.3]{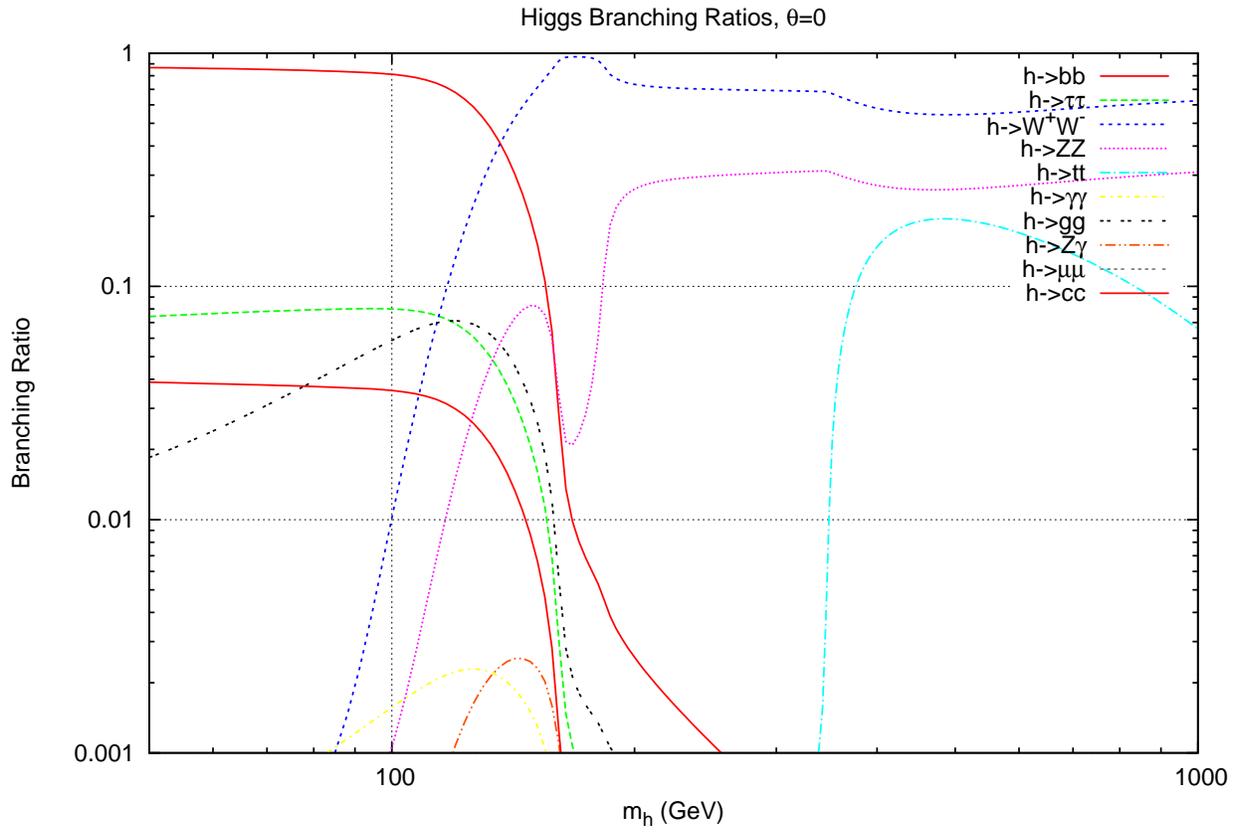}
        \caption{Branching ratio of $h\rightarrow2x$, for $\theta$$=$$0$ and $\alpha$$=$$1$ \cite{hdecay}.}
        \label{h-2x_theta_0}
    \end{center}
\end{figure}

\begin{figure}
    \begin{center}
        \includegraphics[scale=1.3]{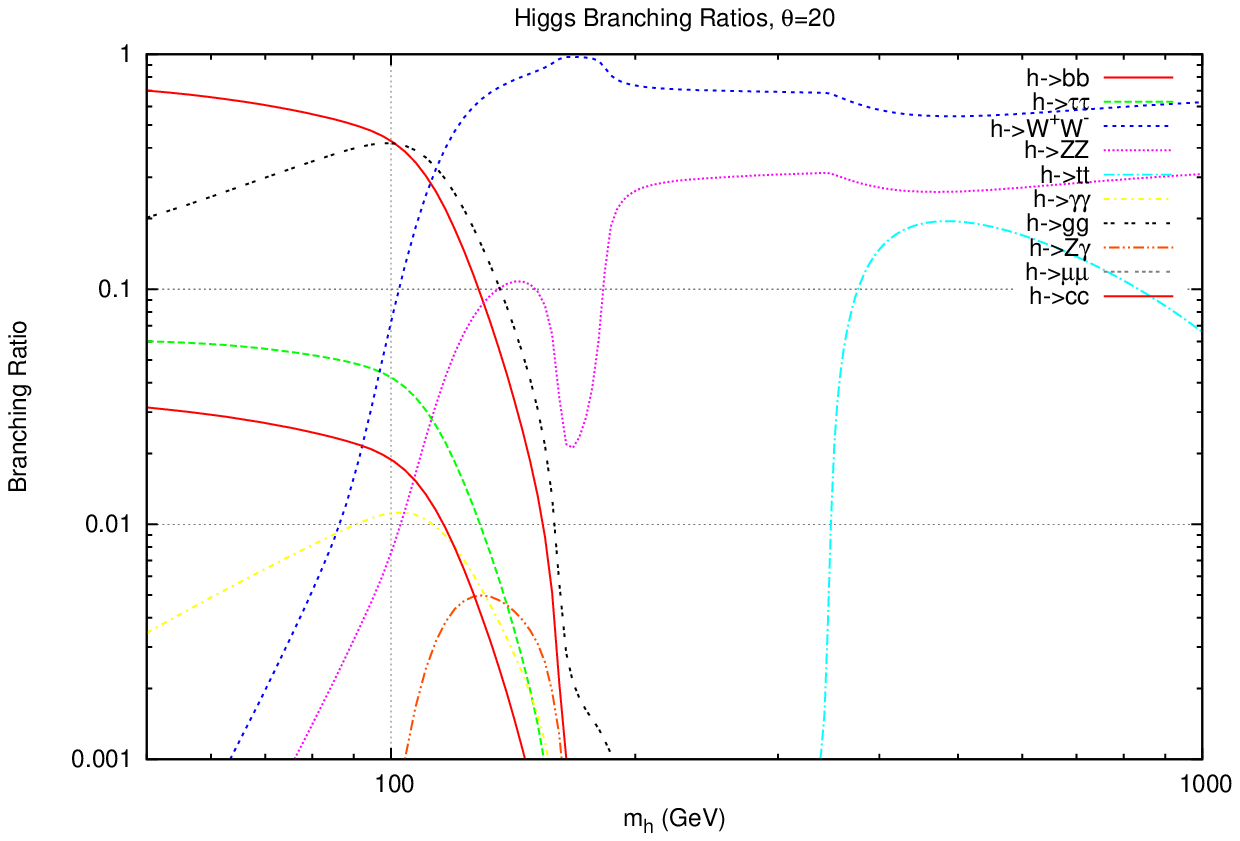}
        \caption{Branching ratio of $h\rightarrow2x$, for $\theta$$=$$20^\circ$ and $\alpha$$=$$1$.}
        \label{h-2x_theta_20}
    \end{center}
\end{figure}

\begin{figure}
    \begin{center}
        \includegraphics[scale=1.3]{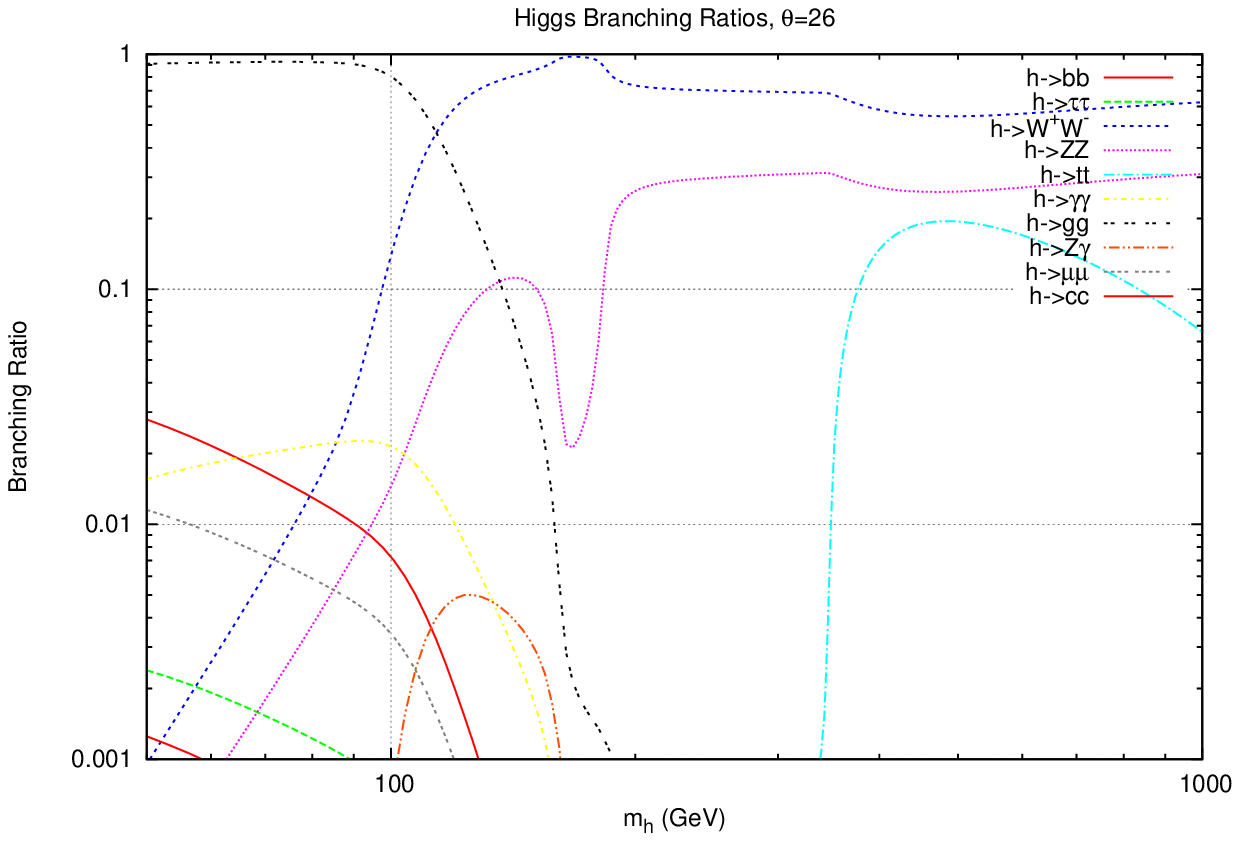}
        \caption{Branching ratio of $h\rightarrow2x$, for $\theta$$=$$26^\circ$ and $\alpha$$=$$1$.}
        \label{h-2x_theta_26}
    \end{center}
\end{figure}

\begin{figure}
    \begin{center}
        \includegraphics[scale=1.3]{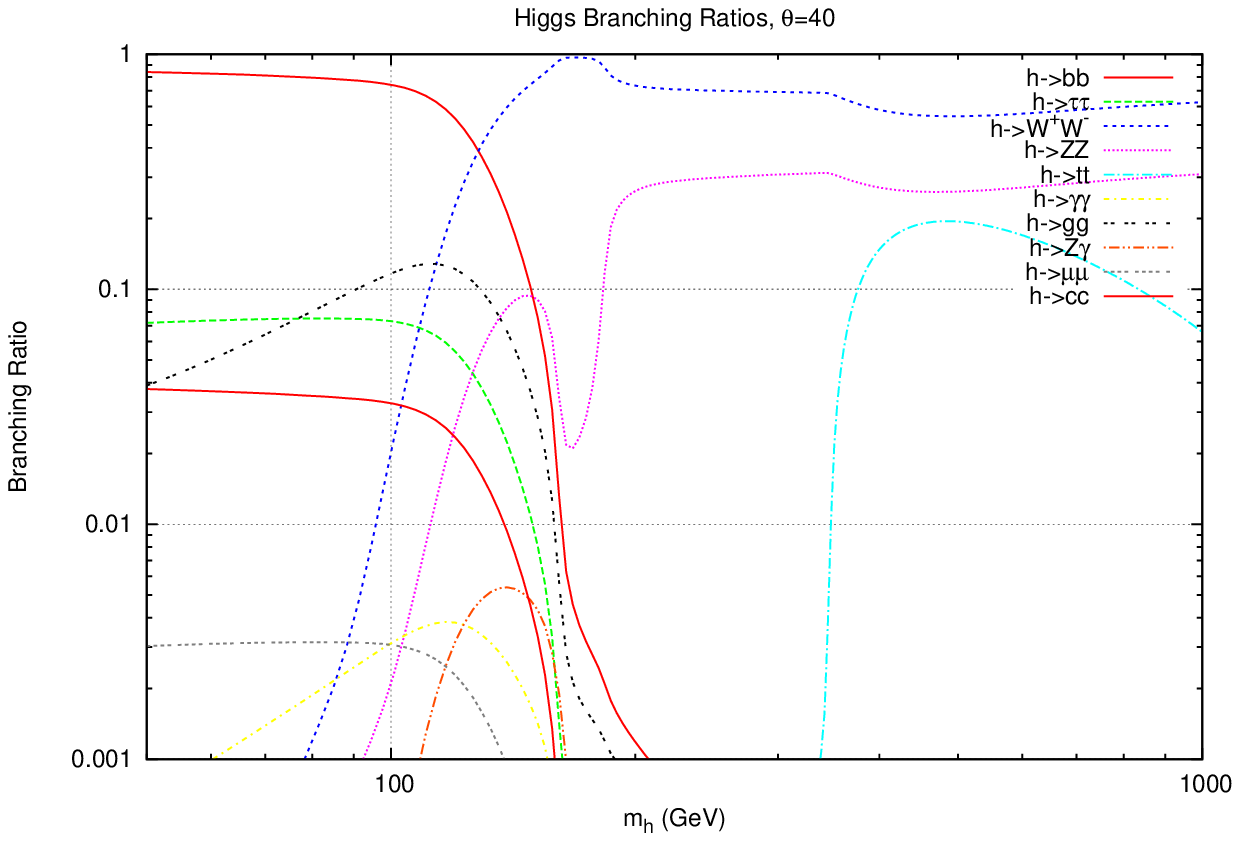}
        \caption{Branching ratio of $h\rightarrow2x$, for $\theta$$=$$40^\circ$ and $\alpha$$=$$1$.}
        \label{h-2x_theta_40}
    \end{center}
\end{figure}

Because of the flavor dependence of the couplings, the branching ratios (BR)
 for $h$ to various final states are altered substantially from those in
 the SM. These branching ratios for $h$ to various final
 states are shown in Figs. \ref{h-2x_theta_0}-\ref{h-2x_theta_40} for values of
 the mixing angle $\theta = 0,~20^\circ,~26^\circ,$ and $40^
 \circ$ respectively.

 For $\theta = 0$, i.e. no mixing, these BR's are the same as for the SM Higgs. Note
 that for both $\theta = 20^\circ$ and $ 26^\circ$, the $gg$ and the $\gamma\gamma$ BR's are enhanced
 substantially compared to the SM.
 This is due to drastic reduction for the $b \bar{b}$ mode from
 an approximate cancellation in the corresponding coupling as can be seen
 from Table 1.
 In particular, for $\theta = 26^\circ$, the effect is quite
 dramatic. For a light Higgs ($m_h$ around $115$ GeV), the usually
 dominant $b \bar{b}$ mode is highly suppressed and the $\gamma\gamma$
 mode is enhanced by a factor of almost $10$ compared to the SM. This is
 to be contrasted with the proposal of Refs. \cite{bn,gl} in which the $h
 \rightarrow \gamma\gamma$ mode is reduced by about a factor of $10$.
 Thus the Higgs signal in this mode for a Higgs mass of $\sim{114 -
 140}$ GeV gets a big enhancement, making its potential
 discovery via this mode much more favorable at the LHC. Such a
 signal may be observable at the Tevatron for a Higgs mass
 $\sim{114}$ as the luminosity accumulates, but would require 
 about 10 fb$^{-1}$ or more of data \cite{tevatron_search}.

 Another interesting effect is the Higgs signal via the $WW^*$  for
 the light Higgs. In the SM, this mode becomes important for the
 Tevatron search for Higgs masses greater than about 135 GeV, where the BR to
 $WW^*$ is approximate equal to that of $b\bar{b}$. Currently
 Tevatron experiments have excluded a SM Higgs with mass around $170$
 GeV (where the BR to $WW^*$ is around $100$ percent) for this mode \cite{teva170}.
 In our framework, for $\theta = 20^\circ$ for example, the crossover
 between the $WW^*$ mode and the $b\bar{b}$ mode takes place
 sooner than $135$ GeV. Thus the Tevatron experiments will be more sensitive to
 the lower mass range than for a SM Higgs, and should be able to exclude masses
 much smaller than $160$ GeV.

 For a heavy Higgs, $m_h > 200$ GeV, the Higgs will be accessible via
 the golden mode $h \rightarrow{ZZ}$. However, in this case, both $h$
 and $s$ decay via this mode with comparable BR's (see Figs. \ref{h-2x_theta_20} and \ref{h-2x_theta_40}
 for $\theta = 20^\circ$ and $40^\circ$). So initially it will be hard to tell
 whether we are seeing $h$ or $s$, a case of Higgs look-alikes. 
 An accurate measurement of this cross
 section times the BR, and the mass of the observed Higgs,
 we will be able to distinguish a heavy $h$ from a heavy $s$, since the production
 cross sections depend on the mixing angle.
 
 \subsection{Top quark physics}\label{sec:4.2}
 In the SM, the $t\rightarrow{c h}$ mode
 is severely suppressed with a BR $\sim10^{-14}$ \cite{tchsm}. In our model, as
 can be seen from eqs.(\ref{SIX}) and (\ref{SEVEN}), although $t\rightarrow{c h}$ is
 zero at tree level, we have a large coupling for
 $t\rightarrow{c s} \sim {2 \epsilon \beta}$ (note $s$ here denotes
the singlet Higgs, not the strange quark).
This gives rise to
 a significant BR  for the $t\rightarrow{c s}$ mode for a Higgs mass
 of up to about $150$ GeV. If the mixing between the h and s  is
 substantial, both decay modes, $t\rightarrow{c s}$ and
 $t\rightarrow{c h}$ will have BR $\sim{10^{-3}}$. With a very large
 $t \bar{t}$ cross section , $\sigma_{t\bar{t}}\sim{10^3}$ pb at the
 LHC, this could be an observable production mode for Higgs bosons at the
 LHC.

 \subsection{$\mathbf{Z^\prime}$ physics} 
\begin{figure}
\begin{center}
       \includegraphics[scale=1.3]{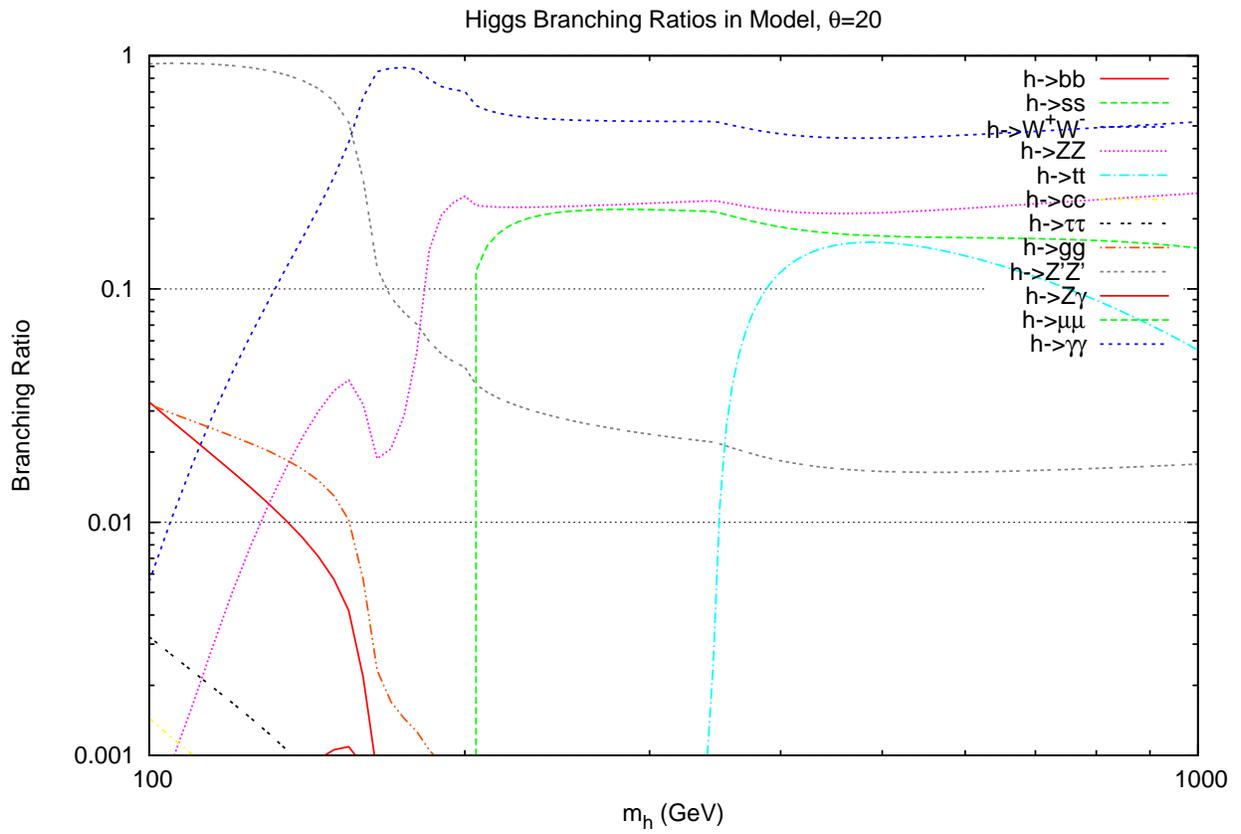}
\caption{Branching ratio of $h\rightarrow2x$ including $h\rightarrow ss$ and $h\rightarrow Z'Z'$
where $m_{Z'}=40$ GeV and $m_{s}=100$ GeV. Here $\theta$$=$$20^\circ$ and $\alpha$$=$$1$.}
\label{h-2x_ZP_theta_20}
\end{center}
\end{figure}

\begin{figure}
\begin{center}
       \includegraphics[scale=1.3]{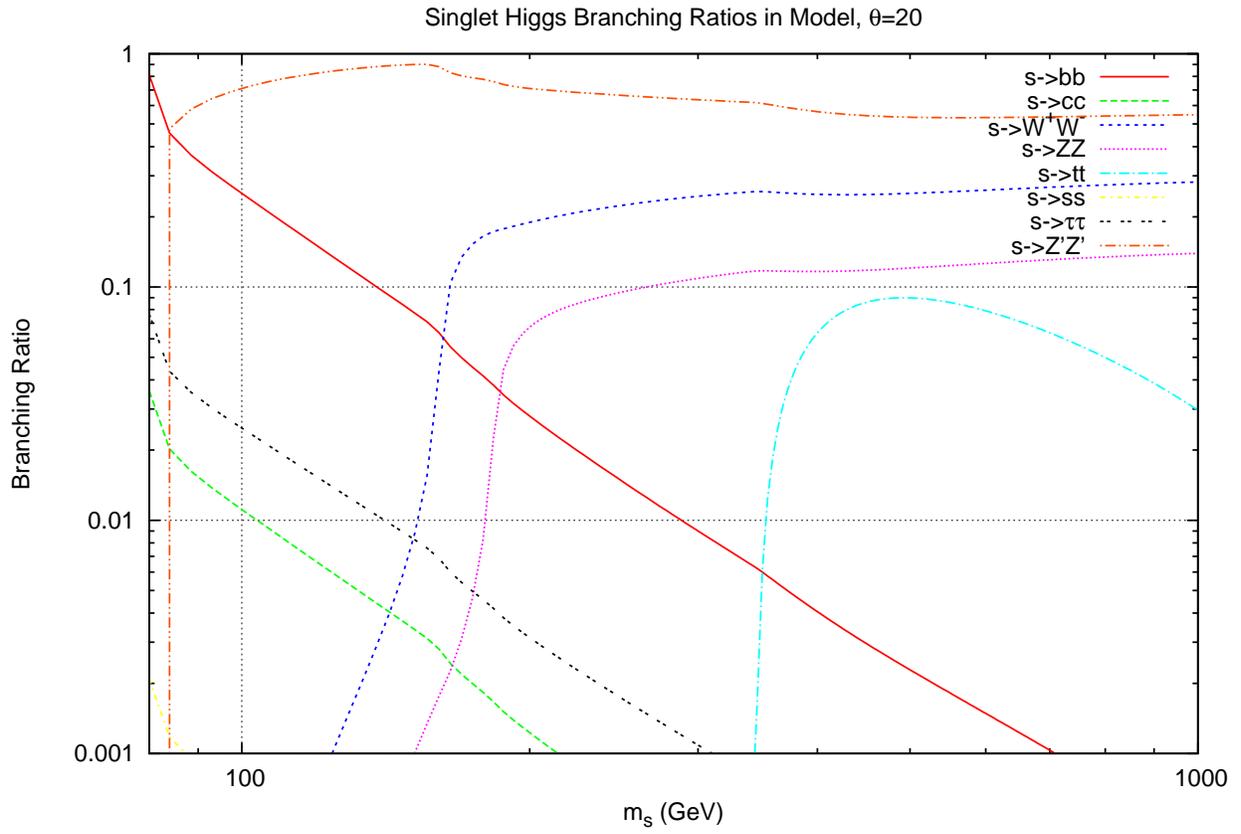}
\caption{Branching ratio of $s\rightarrow2x$ including $s\rightarrow Z'Z'$ where $m_{Z'}=40$ GeV.
Here $\theta$$=$$20^\circ$ and $\alpha$$=$$1$.}
\label{s-2x_ZP_theta_20}
\end{center}
\end{figure}

Our model has a $Z'$ boson near
 the EW scale from the spontaneous breaking of the extra $U(1)$ symmetry.
 As discussed before, since the $Z-Z'$ mixing is very small
 $\sim{10^{-4}}$ or less, its mass is not constrained by the
 very accurately measured $Z$ properties at LEP. Its mass can be as
 low as a few GeV from the existing constraints. This $Z'$ does not
 couple to SM particles with dimension 4 operators. It does
 couple to $s$ at tree level via the $sZ' Z'$ interaction. Thus it
 can be produced via the decay of $s$ (or $h$ if there is a substantial
 mixing between $h$ and $s$). This gives an interesting signal for the
 Higgs decays: $s\rightarrow{Z' Z'}$, $h\rightarrow{Z' Z'}$ if
 allowed kinematically. In Figs. \ref{h-2x_ZP_theta_20} and \ref{s-2x_ZP_theta_20}, we give the BR's
 for $h$ and $s$ decays for a $Z'$ mass of $40$ GeV. The $Z'$ will
 decay to the SM particles via the $Z-Z'$ mixing with the same
 branching ratio as the $Z$. Thus the clear final state signal will
 be $l^+ l^- l^+ l^-$ pairs $(l = e, \mu)$ with each pair having the
 invariant mass of the $Z'$. Such a signal will be easily detectable
 at the LHC. If the $Z'$ happens to be very light, (say a few GeV), and
 the mixing angle is extremely tiny, there is a possibility that the
 $Z'$s may produce displaced vertices at the detector. Both of these
 will be very unconventional signals for Higgs bosons at the
 LHC.

 \subsection{$\mathbf{B_s^0\rightarrow\mu^+\mu^-}$}
 In our framework this decay gets a contribution from an FCNC interaction mediated
 by $s$ exchange.  The amplitude for this decay is
  $A \sim 4 h_{22}^d h_{22}^\ell \epsilon^6 \beta^2$. Taking
   $\beta\sim\epsilon$, $A \sim 4 h_{22}^d h_{22}^\ell \epsilon^8$,
   and with the couplings $h_{22}^d, h_{22}^\ell \sim 1$, we
    obtain the branching ratio, $BR(B_s^0\rightarrow\mu^+\mu^-)\sim 10^{-9}$.
    The current experimental limit for this BR is $4.7\times 10^{-8}$ \cite{pdg}, and
    thus there is a possibility that this decay could be observed at the Tevatron.

 \subsection{Production and decay of heavy fermions} Our framework
 requires vectorlike quarks and leptons, both $SU(2)$ doublets and weak
singlets, with masses around the TeV scale. The heavy quarks
 be pair produced at high energy hadron colliders via the strong
 interaction. For example, for a 1 TeV vectorlike quark, the
 production cross section at the LHC is $\sim{60}$ fb \cite {mangano}. We need
 several such vectorlike quarks for our model. So the total production
 cross section could be as large as a few hundred fb. These will decay to the light
 quarks of the same electric charge  and Higgs bosons ($h$ or $s$):
 Thus the signal will be two high $p_T$
 jets together with the final states arising from the Higgs decays.
 For a heavy Higgs, in the golden mode ($h\rightarrow{Z Z},
 s\rightarrow{Z Z}$, this will give rise to two high $p_T$ jets plus
 four $Z$ bosons. In the case of a light $Z'$, the final state signal
 will be two high $p_T$ jets plus up to 8 charged leptons in the final
 state (with each lepton pair having the invariant mass of the
 $Z'$).

\section {UV Completion}
We present two concrete examples of models from which an effective action like
eq. (\ref{ONE}) can be derived. The first example only reproduces the
second and third generation quark couplings, but its simplicity serves to
introduce the basic issues and mechanisms. The second example is a
complete three generation TeV scale model of quark flavor. The correct lepton couplings
can be obtained from a copy of the same structure used for the down-type quarks.
We assume that neutrino masses benefit from some additional see-saw
mechanism, although it is not obvious that we can't obtain them by
refining the TeV scale flavon model.

\subsection{Two generation model}

For this pedagogical example we will employ two important
simplifications:
\begin{itemize}
\item We only reproduce the second and third generation quark couplings.
In the next subsection we extend this to include the first generation.
\item We will choose charge assignments such that the couplings
$h_{32}^u$, $h_{32}^d$, and $h_{23}^d$ are higher order in $\epsilon$.
As already mentioned nonzero values of these couplings are not needed
to reproduce the observed SM quark masses and mixings.
\end{itemize}

\begin{table}
\begin{center}
    \begin{tabular}{|c|c|c|c|||c|c|c|c|}
    \hline
        \bf{Field} & $\mathbf{U(1)_Y}$ & $\mathbf{U(1)_S}$ & $\mathbf{U(1)_F}$
       &\bf{Field} & $\mathbf{U(1)_Y}$ & $\mathbf{U(1)_S}$ & $\mathbf{U(1)_F}$ \\
    \hline
        $H$      &  1/2  &  0 &  0   &  $U_{1L}$ &  2/3 &  1 &  0  \\
        $S$      &    0  &  1 &  0   &  $U_{1R}$ &  2/3 &  1 &  1  \\
        $F$      &    0  &  0 &  1   &  $U_{2L}$ &  2/3 & -1 &  3  \\
        $q_{3L}$ &  1/6  &  0 &  0   &  $U_{2R}$ &  2/3 & -1 &  3  \\
        $q_{2L}$ &  1/6  &  0 &  2   &  $D_{1L}$ & -1/3 & -1 & -1  \\
        $u_{3R}$ &  2/3  &  0 &  0   &  $D_{1R}$ & -1/3 & -1 & -1  \\
        $u_{2R}$ &  2/3  &  0 &  3   &  $D_{2L}$ & -1/3 &  2 &  3  \\
        $d_{3R}$ & -1/3  &  0 & -1   &  $D_{2R}$ & -1/3 &  2 &  2  \\
        $d_{2R}$ & -1/3  &  0 &  3   &  $D_{3L}$ & -1/3 &  1 &  3  \\
        $Q_{1L}$ &  1/6  & -1 & -1   &  $D_{3R}$ & -1/3 &  1 &  3  \\
        $Q_{1R}$ &  1/6  & -1 &  0   &           &      &    &     \\
        $Q_{2L}$ &  1/6  &  1 &  1   &           &      &    &     \\
        $Q_{2R}$ &  1/6  &  1 &  2   &           &      &    &     \\
        $Q_{3L}$ &  1/6  & -1 &  3   &           &      &    &     \\
        $Q_{3R}$ &  1/6  & -1 &  2   &           &      &    &     \\
        $Q_{4L}$ &  1/6  &  2 &  2   &           &      &    &     \\
        $Q_{4R}$ &  1/6  &  2 &  1   &           &      &    &     \\
    \hline
    \end{tabular}
\end{center}
\caption{\label{table:smallcharge} Charge assignments in the two generation model for the scalar fields $H$, $S$, $F$, and the
SM quark fields $q_{3L}$, $q_{2L}$, $u_{3R}$, $u_{2R}$, $d_{3R}$, and $d_{2R}$.
Also listed are the color triplet weak doublet
heavy quark pairs $Q_{iL}$, $Q_{iR}$ and
the color triplet weak singlet heavy quark pairs
$U_{iL}$, $U_{iR}$, $D_{iL}$, $D_{iR}$.}
\end{table}

With these simplifications we postulate a TeV scale model with
the field content shown in Table \ref{table:smallcharge}, where the hypercharges are
listed along with the charge assignments under $U(1)_S$ and $U(1)_F$.
The Higgs doublet $H$ is the only scalar that carries hypercharge,
while the SM singlet $S$ is the only scalar carrying $U(1)_S$ charge.
The SM singlet flavon $F$ is the only scalar carrying $U(1)_F$
charge. The SM quarks are neutral under $U(1)_S$. The third generation
up-type quark fields also carry no $U(1)_F$ charge, while the other
quark fields have flavor-dependent nonzero $U(1)_F$ charges.

We introduce four pairs of new color triplet weak doublet fermion fields
$Q_{iL}$, $Q_{iR}$, two pairs of color triplet up-type weak singlets $U_{iL}$, $U_{iR}$,
and three pairs of color triplet down-type weak singlets
$D_{iL}$, $D_{iR}$. Each pair is vectorlike with respect to
the SM gauge group and $U(1)_S$, thus no anomalies are introduced with
respect to these gauge groups, and each vectorlike pair naturally
acquires a Dirac mass of order $M$ (when they have the same $U(1)_F$ charge)
or of order the vev of $F$ (when their $U(1)_F$ charges differ by one).
We assume that both the vev of $F$ and $M$ are greater than, but of order of, a TeV.
Any residual anomaly in $U(1)_F$ can be handled either by introducing
more heavy fermions or using the Green-Schwarz mechanism above the TeV scale.

With these charge assignments the only dimension 4 couplings
involving the second and third generation SM quarks are:
\be
&&\hspace{-10pt}
  f_1\overline{q}_{3L}u_{3R}\bar{H}
+ f_2\overline{q}_{3L} Q_{1R} S
+ f_3\overline{D}_{1L} d_{3R} S^\dagger
+ f_4\overline{q}_{2L} Q_{2R} S^\dagger
\nonumber\\
&&\hspace{-10pt}
+ f_5\overline{U}_{1L} u_{3R} S
+ f_6\overline{q}_{2L} Q_{3R} S
+ f_7\overline{U}_{2L} u_{2R} S^\dagger
+ f_8\overline{D}_{3L} d_{2R} S
+ h.c. \quad ,
\label{FOURTEEN}
\ee
where the $f_i$ are dimensionless coupling constants.
Thus the top quark receives the correct mass from electroweak
symmetry breaking for $\vert f_1 \vert \simeq 1$. The other
couplings involve the $S$ scalar, but not the Higgs $H$ or
the flavon $F$.
Both electroweak symmetry breaking and flavor symmetry breaking
are communicated to the rest of the SM quark sector via a
Froggart-Nielsen type mechanism, integrating out the heavy
TeV scale fermions from tree level diagrams that connect
SM quark left doublets to SM quark right singlets and to
$H$ or $\bar{H}$.

The renormalizable couplings involving just the
heavy fermions are:
\be
&&\hspace{-10pt}
  f_9   \overline{Q}_{1R} Q_{1L} F
+ f_{10}\overline{Q}_{1L} D_{1R} H
+ M\overline{D}_{1R} D_{1L}
\nonumber\\
&&\hspace{-10pt}
+ f_{11}\overline{Q}_{2R} Q_{2L} F
+ f_{12}\overline{Q}_{2L} U_{1R} \bar{H}
+ f_{13}\overline{U}_{1R} U_{1L} F
\\
&&\hspace{-10pt}
+ f_{14}\overline{Q}_{3R} Q_{3L} F^\dagger
+ f_{15}\overline{Q}_{3L} U_{2R} \bar{H}
+ M\overline{U}_{2R} U_{2L}
\nonumber\\
&&\hspace{-10pt}
+ f_{16}\overline{Q}_{2L} Q_{4R} S^\dagger
+ f_{17}\overline{Q}_{4L} Q_{2R} S
+ f_{18}\overline{Q}_{4R} Q_{4L} F^\dagger
+ f_{19}\overline{Q}_{4L} D_{2R} H
\nonumber\\
&&\hspace{-10pt}
+ f_{20}\overline{D}_{2R} D_{2L} F^\dagger
+ f_{21}\overline{D}_{2L} D_{3R} S
+ M\overline{D}_{3L} D_{3R}
+ h.c. \quad .
\nonumber
\ee

Thus, integrating out the heavy fermions in the
tree level diagram composed from the couplings
\be
  f_2\overline{q}_{3L} Q_{1R} S
+ f_9   \overline{Q}_{1R} Q_{1L} F
+ f_{10}\overline{Q}_{1L} D_{1R} H
+ M\overline{D}_{1R} D_{1L}
+ f_3\overline{D}_{1L} d_{3R} S^\dagger
\label{twenty}
\ee
produces an effective coupling below the
TeV scale proportional to
\be
f_2f_3f_9f_{10}\frac{F}{M}\frac{S^\dagger S}{M^2}\overline{q}_{3L}d_{3R}H + h.c.
\; .
\ee

Integrating out the heavy fermions in the
tree level diagram composed from the couplings
\be
f_4\overline{q}_{2L} Q_{2R} S^\dagger
+ f_{11}\overline{Q}_{2R} Q_{2L} F
+ f_{12}\overline{Q}_{2L} U_{1R} \bar{H}
+ f_{13}\overline{U}_{1R} U_{1L} F
+ f_5\overline{U}_{1L} u_{3R} S
\label{twentytwo}
\ee
produces an effective coupling below the
TeV scale proportional to
\be
f_4f_5f_{11}f_{12}f_{13}\frac{F^2}{M^2}\frac{S^\dagger S}{M^2}\overline{q}_{2L}u_{3R}\bar{H} + h.c.
\; .
\ee

Integrating out the heavy fermions in the
tree level diagram composed from the couplings
\be
f_6\overline{q}_{2L} Q_{3R} S
+ f_{14}\overline{Q}_{3R} Q_{3L} F^\dagger
+ f_{15}\overline{Q}_{3L} U_{2R} \bar{H}
+ M\overline{U}_{2R} U_{2L}
+ f_7\overline{U}_{2L} u_{2R} S^\dagger
\label{twentyfour}
\ee
produces an effective coupling below the
TeV scale proportional to
\be
f_6f_7f_{14}f_{15}\frac{F^\dagger}{M}\frac{S^\dagger S}{M^2}\overline{q}_{2L}u_{2R}\bar{H}
+ h.c.\; .
\ee

Finally, integrating out the heavy fermions in the
tree level diagram composed from the couplings
\be
&&\hspace{-10pt}
f_4\overline{q}_{2L} Q_{2R} S^\dagger
+ f_{17}^*\overline{Q}_{2R} Q_{4L} S^\dagger
+ f_{19}\overline{Q}_{4L} D_{2R} H
\nonumber\\
&&\hspace{-10pt}
+ f_{20}\overline{D}_{2R} D_{2L} F^\dagger
+ f_{21}\overline{D}_{2L} D_{3R} S
+ M\overline{D}_{3R} D_{3L}
+ f_8\overline{D}_{3L} d_{2R} S
\label{twentysixfirst}
\ee
produces an effective coupling below the
TeV scale proportional to
\be
f_4f_8f_{17}^*f_{19}f_{20}f_{21}\frac{F^\dagger}{M}\frac{(S^\dagger S)^2}{M^4}
\overline{q}_{2L}d_{2R}H + h.c.
\; .
\ee
There is an additional very similar tree level diagram contributing to $h_{22}^d$ composed
from the couplings
\be
&&\hspace{-10pt}
f_4\overline{q}_{2L} Q_{2R} S^\dagger
+ f_{11}\overline{Q}_{2R} Q_{2L} F
+ f_{16}\overline{Q}_{2L} Q_{4R} S^\dagger
+ f_{18}\overline{Q}_{4R} Q_{4L} F^\dagger
+ f_{19}\overline{Q}_{4L} D_{2R} H
\nonumber\\
&&\hspace{-10pt}
+ f_{20}\overline{D}_{2R} D_{2L} F^\dagger
+ f_{21}\overline{D}_{2L} D_{3R} S
+ M\overline{D}_{3R} D_{3L}
+ f_8\overline{D}_{3L} d_{2R} S
\label{twentysixsecond}
\ee
which produces an effective coupling below the
TeV scale proportional to
\be
f_4f_8f_{11}f_{16}f_{18}f_{19}f_{20}f_{21}\frac{F(F^\dagger)^2}{M^3}\frac{(S^\dagger S)^2}{M^4}
\overline{q}_{2L}d_{2R}H + h.c.
\; .
\ee

\begin{figure}
    \centering
        \includegraphics[scale=0.7]{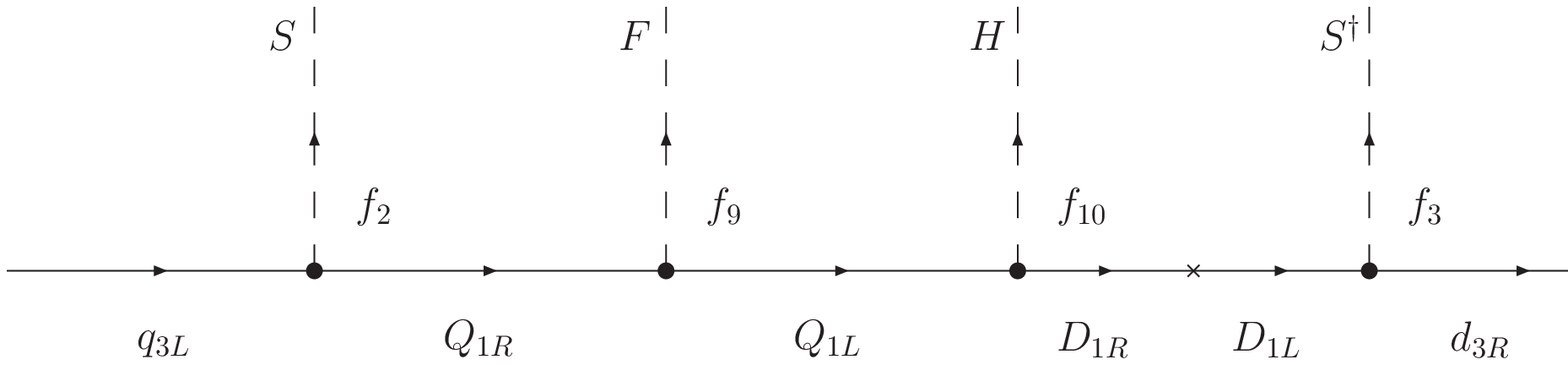}
    \caption{The Feynman diagram associated with eq. (\ref{twenty})}
    \label{fig:diag20}
\end{figure}

\begin{figure}
    \centering
        \includegraphics[scale=0.7]{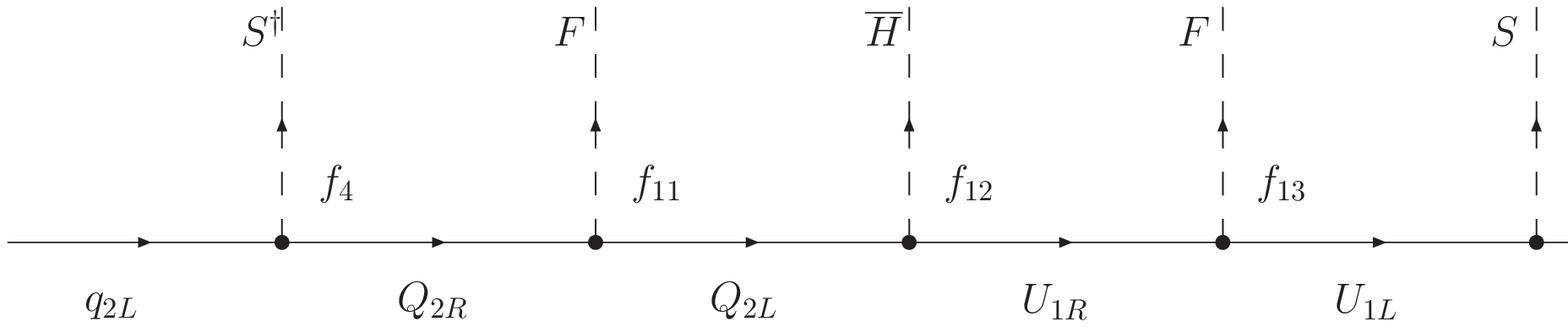}
    \caption{The Feynman diagram associated with eq. (\ref{twentytwo})}
    \label{fig:diag22}
\end{figure}

\begin{figure}
    \centering
        \includegraphics[scale=0.7]{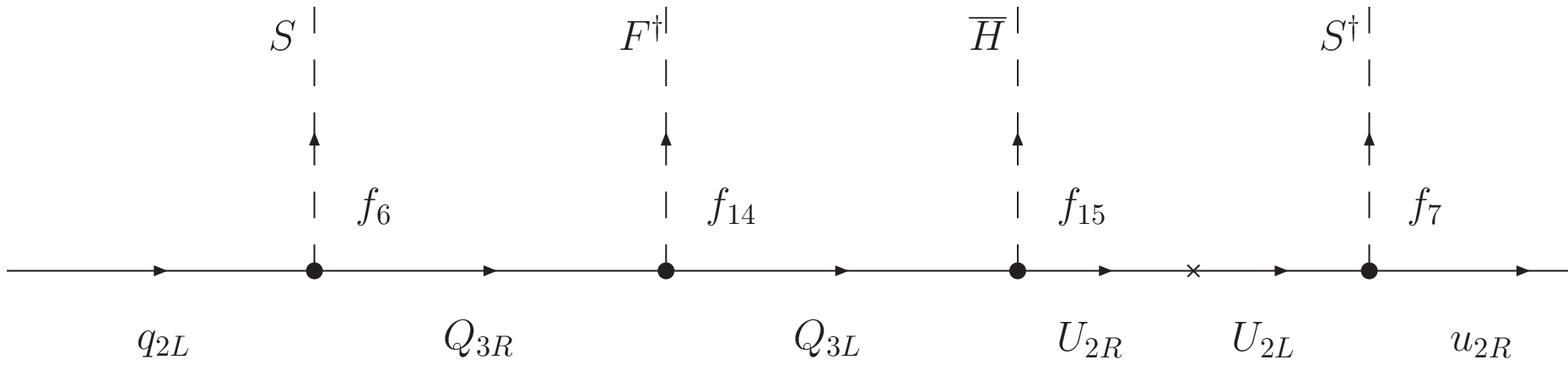}
    \caption{The Feynman diagram associated with eq. (\ref{twentyfour})}
    \label{fig:diag24}
\end{figure}

\begin{figure}
    \centering
        \includegraphics[scale=0.7]{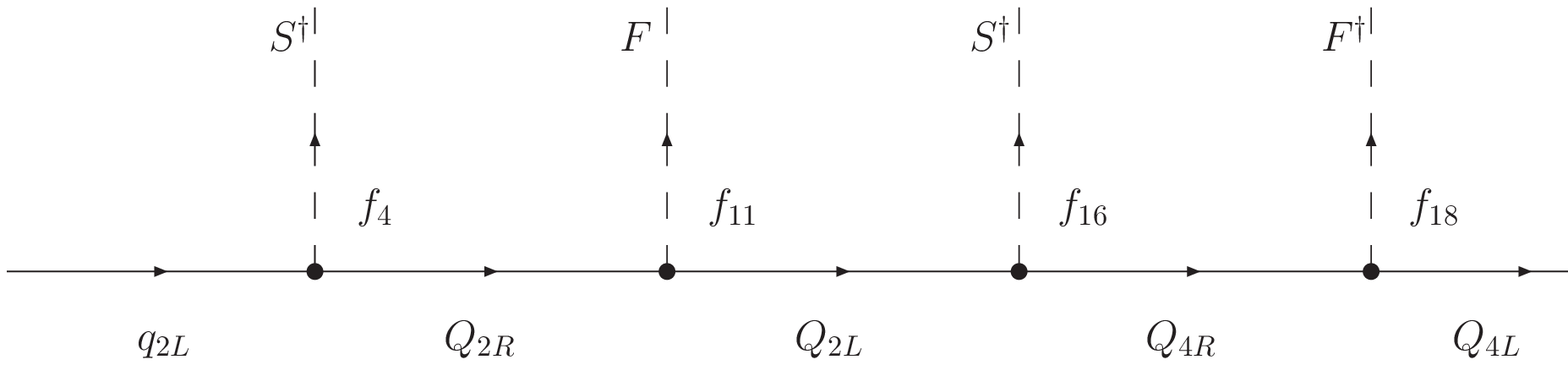}
\end{figure}
\begin{figure}
    \centering
        \includegraphics[scale=0.7]{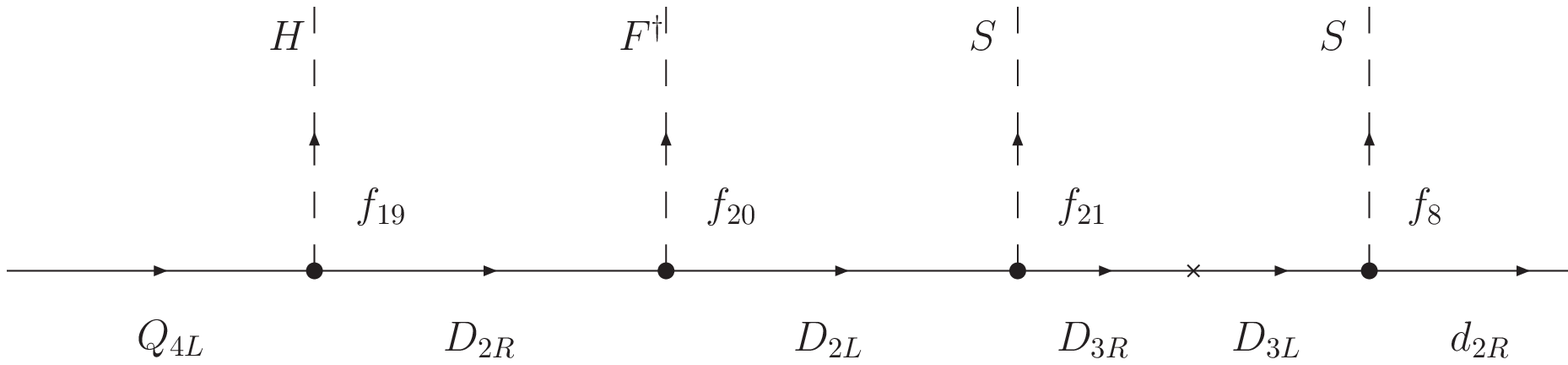}
    \caption{The Feynman diagram associated with eq. (\ref{twentysixsecond})}
    \label{fig:diag26}
\end{figure}

\subsection{Three generation model}

Here we present an concrete example of a full three generation TeV scale model that
reproduces an effective action like eq. (\ref{ONE}) at the electroweak scale.
This model uses a single electroweak messenger scalar $S$, but employs three TeV scale flavon scalars
$F_1$, $F_2$, and $F_3$, each corresponding to a different broken $U(1)_{F_i}$ flavon symmetry.
As before the SM quarks are neutral under $U(1)_S$. The third generation
up-type quark fields also carry no $U(1)_{F_i}$ charges, while the other
quark fields have flavor-dependent nonzero $U(1)_{F_i}$ charges.

The model has a rather large number of new heavy fermions:
seven pairs of new color triplet weak doublet fermion fields
$Q_{iL}$, $Q_{iR}$, six pairs of color triplet up-type weak singlets $U_{iL}$, $U_{iR}$,
and eight pairs of color triplet down-type weak singlets
$D_{iL}$, $D_{iR}$. Each pair is vectorlike with respect to
the SM gauge group and $U(1)_S$, thus no anomalies are introduced with
respect to these gauge groups, and each vectorlike pair naturally
acquires a Dirac mass of order $M$ (when they have the same $U(1)_{F_i}$ charges)
or of order the vev of some $F_i$ (when one of their $U(1)_{F_i}$ charges differs by one).
We assume that both the $F_i$ vevs and $M$ are of order a TeV.
Any residual anomaly in the $U(1)_{F_i}$ symmetries can be handled either by introducing
more heavy fermions or using the Green-Schwarz mechanism at the TeV scale.

We do not suggest that this model is the most efficient one implementing
the basic concepts of our proposal. We have made an explicit trade-off, in
some sense, of maximizing the number of the new heavy fermions required in order
to minimize the complexity of the messenger sector and the charge assignments.

\begin{table}
\begin{center}
    \begin{tabular}{|c|c|c|c|c|c|||c|c|c|c|c|c|}
    \hline
        \bf{Field} & $\mathbf{U(1)_Y}$ & $\mathbf{U(1)_S}$ & $\mathbf{U(1)_{F1}}$ & $\mathbf{U(1)_{F2}}$ & $\mathbf{U(1)_{F3}}$
       &\bf{Field} & $\mathbf{U(1)_Y}$ & $\mathbf{U(1)_S}$ & $\mathbf{U(1)_{F1}}$ & $\mathbf{U(1)_{F2}}$ & $\mathbf{U(1)_{F3}}$ \\
    \hline
$q_{1L}$ & 1/6 & 0 & 1 & 2 & 1 & $U_{1L}$ & 2/3 & 1 & 0 & 1 & 1 \\
$q_{2L}$ & 1/6 & 0 & 0 & 1 & 0 & $U_{1R}$ & 2/3 & 1 & 0 & 1 & 0 \\
$q_{3L}$ & 1/6 & 0 & 0 & 0 & 0 & $U_{2L}$ & 2/3 & 1 & 1 & 0 & 0 \\
$u_{1R}$ & 2/3 & 0 & 0 & 1 & 1 & $U_{2R}$ & 2/3 & 1 & 1 & 1 & 0 \\
$u_{2R}$ & 2/3 & 0 & 1 & 0 & 0 & $U_{3L}$ & 2/3 & 2 & 2 & 1 & -1 \\
$u_{3R}$ & 2/3 & 0 & 0 & 0 & 0 & $U_{3R}$ & 2/3 & 2 & 2 & 1 & -1 \\
$d_{1R}$ & -1/3 & 0 & 1 & 2 & 0 & $U_{4L}$ & 2/3 & 2 & 1 & 1 & 0 \\
$d_{2R}$ & -1/3 & 0 & 0 & 1 & 1 & $U_{4R}$ & 2/3 & 2 & 2 & 1 & 0 \\
$d_{3R}$ & -1/3 & 0 & 1 & 0 & 0 & $U_{5L}$ & 2/3 & 2 & 0 & 1 & 0 \\
$Q_{1L}$ & 1/6 & 1 & 1 & 2 & 0 & $U_{5R}$ & 2/3 & 2 & 0 & 2 & 0 \\
$Q_{1R}$ & 1/6 & 1 & 1 & 2 & 1 & $U_{6L}$ & 2/3 & 3 & 0 & 2 & 0 \\
$Q_{2L}$ & 1/6 & 1 & 1 & 1 & 0 & $U_{6R}$ & 2/3 & 3 & 1 & 2 & 0 \\
$Q_{2R}$ & 1/6 & 1 & 0 & 1 & 0 & $D_{1L}$ & -1/3 & 1 & 1 & 2 & 0 \\
$Q_{3L}$ & 1/6 & 1 & 1 & 0 & 0 & $D_{1R}$ & -1/3 & 1 & 1 & 2 & 1 \\
$Q_{3R}$ & 1/6 & 1 & 0 & 0 & 0 & $D_{2L}$ & -1/3 & 1 & 0 & 1 & 1 \\
$Q_{4L}$ & 1/6 & 2 & 1 & 0 & 0 & $D_{2R}$ & -1/3 & 1 & 1 & 1 & 1 \\
$Q_{4R}$ & 1/6 & 2 & 1 & 1 & 0 & $D_{3L}$ & -1/3 & 1 & 1 & 0 & 0 \\
$Q_{5L}$ & 1/6 & 2 & 2 & 2 & 0 & $D_{3R}$ & -1/3 & 1 & 1 & 0 & 0 \\
$Q_{5R}$ & 1/6 & 2 & 1 & 2 & 0 & $D_{4L}$ & -1/3 & 2 & 1 & 0 & 0 \\
$Q_{6L}$ & 1/6 & 2 & 2 & 1 & -1 & $D_{4R}$ & -1/3 & 2 & 1 & 0 & 0 \\
$Q_{6R}$ & 1/6 & 2 & 2 & 1 & 0 & $D_{5L}$ & -1/3 & 2 & 1 & 1 & 1 \\
$Q_{7L}$ & 1/6 & 3 & 1 & 2 & 0 & $D_{5R}$ & -1/3 & 2 & 1 & 1 & 0 \\
$Q_{7R}$ & 1/6 & 3 & 2 & 2 & 0 & $D_{6L}$ & -1/3 & 2 & 1 & 2 & 1 \\
$H$      & 1/2 & 0 & 0 & 0 & 0 & $D_{6R}$ & -1/3 & 2 & 1 & 2 & 0 \\
$S$      & 0 & 1 & 0 & 0 & 0 & $D_{7L}$ & -1/3 & 3 & 1 & 1 & 0 \\
$F_1$    & 0 & 0 & 1 & 0 & 0 & $D_{7R}$ & -1/3 & 3 & 1 & 2 & 0 \\
$F_2$    & 0 & 0 & 0 & 1 & 0 & $D_{8L}$ & -1/3 & 3 & 1 & 2 & 0 \\
$F_3$    & 0 & 0 & 0 & 0 & 1 & $D_{8R}$ & -1/3 & 3 & 1 & 2 & 1 \\
    \hline
    \end{tabular}
\end{center}
\caption{\label{table:bigcharge} Charge assignments in the three generation model for the scalar fields $H$, $S$, $F_i$, the
SM quark fields $q_{iL}$, $u_{iR}$, $d_{iR}$, and the heavy
quark pairs $Q_{iL}$, $Q_{iR}$, $U_{iL}$, $U_{iR}$, $D_{iL}$, $D_{iR}$.}
\end{table}

With the charge assignments listed in Table \ref{table:bigcharge} the only dimension 4 couplings of fermions
to the Higgs scalar are
\be
&&\hspace*{-20pt}
f_1\overline{q}_{3L}u_{3R}\bar{H}
+f_2\overline{Q}_{2L}U_{2R}\bar{H}
+f_3\overline{Q}_{4R}U_{4L}\bar{H}
+f_4\overline{Q}_{6L}U_{3R}\bar{H} \nonumber\\
&&\hspace*{-20pt}
+f_5\overline{Q}_{7L}U_{6R}\bar{H}
+f_6\overline{Q}_{3L}D_{3R}H
+f_7\overline{Q}_{4L}D_{4R}H
+f_8\overline{Q}_{7L}D_{7R}H + h.c. \; .
\ee

The only dimension 4 couplings of fermions to the the $S$ messenger
scalar are
\be
&&\hspace*{-20pt}
f_9\overline{q}_{1L}Q_{1R} S^\dagger
+f_{10}\overline{q}_{2L}Q_{2R} S^\dagger
+f_{11}\overline{q}_{3L}Q_{3R} S^\dagger
+f_{12}\overline{U}_{1L}u_{1R} S  \nonumber\\
&&\hspace*{-20pt}
+f_{13}\overline{U}_{2L}u_{2R} S
+f_{14}\overline{D}_{1L}d_{1R} S
+f_{15}\overline{D}_{2L}d_{2R} S
+f_{16}\overline{D}_{3L}d_{3R} S  \nonumber\\
&&\hspace*{-20pt}
+f_{17}\overline{Q}_{2L}Q_{4R} S^\dagger
+f_{18}\overline{Q}_{1L}Q_{5R} S^\dagger
+f_{19}\overline{Q}_{7L}Q_{5R} S^\dagger
+f_{20}\overline{Q}_{5L}Q_{7R} S^\dagger  \nonumber\\
&&\hspace*{-20pt}
+f_{21}\overline{U}_{4L}U_{2R} S
+f_{22}\overline{U}_{5L}U_{1R} S
+f_{23}\overline{U}_{6L}U_{5R} S
+f_{24}\overline{D}_{4L}D_{3R} S  \\
&&\hspace*{-20pt}
+f_{25}\overline{D}_{3L}D_{4R} S^\dagger
+f_{26}\overline{D}_{5L}D_{2R} S
+f_{27}\overline{D}_{6L}D_{1R} S
+f_{28}\overline{D}_{1L}D_{6R} S^\dagger  \nonumber\\
&&\hspace*{-20pt}
+f_{29}\overline{D}_{7L}D_{5R} S
+f_{30}\overline{D}_{8L}D_{6R} S
+f_{31}\overline{D}_{6L}D_{8R} S^\dagger + h.c. \;. \nonumber
\ee

The direct fermion mass terms and mixings consistent
with the flavon symmetries and SM gauge symmetries
generated by operators of dimension 4 or less are
\be
&&\hspace*{-20pt}
f_{32}\overline{Q}_{1L} Q_{1R} F_3^\dagger
+f_{33}\overline{Q}_{2L} Q_{2R} F_1
+f_{34}\overline{Q}_{3L} Q_{3R} F_1
+f_{35}\overline{Q}_{3L} Q_{3R} F_2^\dagger  \nonumber\\
&&\hspace*{-20pt}
+f_{36}\overline{Q}_{4L} Q_{4R} F_2^\dagger
+f_{37}\overline{Q}_{5L} Q_{5R} F_1
+f_{38}\overline{Q}_{5L} Q_{6R} F_2
+f_{39}\overline{Q}_{6L} Q_{6R} F_3^\dagger  \nonumber\\
&&\hspace*{-20pt}
+f_{40}\overline{Q}_{7L} Q_{7R} F_1^\dagger
+f_{41}\overline{U}_{1L} U_{1R} F_3
+f_{42}\overline{U}_{2L} U_{2R} F_2^\dagger
+M\overline{U}_{3L} U_{3R}
+f_{43}\overline{U}_{3L} U_{4R} F_3^\dagger  \nonumber\\
&&\hspace*{-20pt}
+f_{44}\overline{U}_{4L} U_{4R} F_1^\dagger
+f_{45}\overline{U}_{5L} U_{5R} F_2^\dagger
+f_{46}\overline{U}_{6L} U_{6R} F_1^\dagger
+f_{47}\overline{D}_{1L} D_{1R} F_3^\dagger  \\
&&\hspace*{-20pt}
+f_{48}\overline{D}_{2L} D_{2R} F_1^\dagger
+M\overline{D}_{3L} D_{3R}
+M\overline{D}_{4L} D_{4R}
+f_{49}\overline{D}_{5L} D_{5R} F_3         \nonumber\\
&&\hspace*{-20pt}
+f_{50}\overline{D}_{4L} D_{5R} F_2^\dagger
+f_{51}\overline{D}_{6L} D_{6R} F_3
+f_{52}\overline{D}_{7L} D_{7R} F_2^\dagger
+M\overline{D}_{8L} D_{7R}
+f_{53}\overline{D}_{8L} D_{8R} F_3^\dagger + h.c. \;, \nonumber
\ee
where for simplicity of notation we have used $M$ to denote
all the TeV scale mass parameters.

Thus, integrating out the heavy fermions in the
tree level diagram composed from the couplings
\be
  f_{11}\overline{q}_{3L} Q_{3R} S^\dagger
+ f_{34}^*\overline{Q}_{3R} Q_{3L} F_1^\dagger
+ f_{10}\overline{Q}_{3L} D_{3R} H
+ M\overline{D}_{3R} D_{3L}
+ f_3\overline{D}_{3L} d_{3R} S
\label{twentyb}
\ee
produces an effective coupling below the
TeV scale proportional to
\be
f_{11}f_{34}^*f_{10}f_{3}\frac{F_1^\dagger}{M}\frac{S^\dagger S}{M^2}\overline{q}_{3L}d_{3R}H + h.c.
\; .
\ee

Integrating out the heavy fermions in the tree level diagram
composed from the couplings
\be
f_{10}\overline{q}_{2L} Q_{2R} S^\dagger
+ f_{33}^*\overline{Q}_{2R} Q_{2L} F_1^\dagger
+ f_{2}\overline{Q}_{2L} U_{2R} \bar{H}
+ f_{42}^*\overline{U}_{2R} U_{2L} F_2
+ f_{13}\overline{U}_{2L} u_{2R} S
\label{twentyoneb}
\ee
produces an effective coupling below the
TeV scale proportional to
\be
f_{10}f_{33}^*f_{2}f_{42}^*f_{13}\frac{F_1^\dagger F_2}{M^2}\frac{S^\dagger S}{M^2}\overline{q}_{2L}u_{2R}\bar{H} + h.c.
\; .
\ee

Integrating out the heavy fermions in the tree level diagram
composed from the couplings
\be
&&\hspace*{-20pt}
f_{10}\overline{q}_{2L} Q_{2R} S^\dagger
+ f_{33}^*\overline{Q}_{2R} Q_{2L} F_1^\dagger
+f_{17}\overline{Q}_{2L} Q_{4R} S^\dagger
+f_{36}^*\overline{Q}_{4R} Q_{4L} F2
+ f_{7}\overline{Q}_{4L} D_{4R} H \nonumber\\
&&\hspace*{-20pt}
+ M\overline{D}_{4R} D_{4L}
+ f_{24}^*\overline{D}_{4L} D_{3R} S
+ M\overline{D}_{3R} D_{3L}
+ f_3\overline{D}_{3L} d_{3R} S
\label{twentytwob}
\ee
produces an effective coupling below the
TeV scale proportional to
\be
f_{10}f_{33}^*f_{17}f_{36}^*f_{7}f_{24}^*f_{3}\frac{F_1^\dagger F_2}{M^2}\frac{(S^\dagger S)^2}{M^4}\overline{q}_{2L}d_{3R}H
+ h.c.\; .
\ee

Integrating out the heavy fermions in the tree level diagram
composed from the couplings
\be
&&\hspace*{-20pt}
  f_{11}\overline{q}_{3L} Q_{3R} S^\dagger
+ f_{34}^*\overline{Q}_{3R} Q_{3L} F_1^\dagger
+ f_{10}\overline{Q}_{3L} D_{3R} H
+ f_{24}^*\overline{D}_{3R} D_{4L} S^\dagger
+ f_{50}\overline{D}_{4L} D_{5R} F_2^\dagger \nonumber\\
&&\hspace*{-20pt}
+ f_{49}^*\overline{D}_{5R} D_{5L} F_3^\dagger
+ f_{26}\overline{D}_{5L} D_{2R} S
+ f_{48}^*\overline{D}_{2R} D_{2L} F_1
+ f_{15}\overline{D}_{2L} d_{2R} S
\label{twentytwoc}
\ee
produces an effective coupling below the
TeV scale proportional to
\be
f_{11}f_{34}^*f_{10}f_{24}^*f_{50}f_{49}^*f_{26}f_{48}^*f_{15}\frac{F_1^\dagger F_1 F_2^\dagger F_3^\dagger}{M^4}
\frac{(S^\dagger S)^2}{M^4}\overline{q}_{3L}d_{2R}H
+ h.c.\; .
\ee
There is also another tree level contribution to $h_{32}^d$, proportional to
\be
f_{11}f_{34}^*f_{10}f_{24}f_{50}f_{49}^*f_{26}f_{48}^*f_{15}\frac{F_1^\dagger F_1 F_2^\dagger F_3^\dagger}{M^4}
\frac{(S^\dagger S)^2}{M^4}\overline{q}_{3L}d_{2R}H
+ h.c.\; .
\ee

Integrating out the heavy fermions in the tree level diagram
composed from the couplings
\be
&&\hspace*{-20pt}
f_{10}\overline{q}_{2L} Q_{2R} S^\dagger
+ f_{33}^*\overline{Q}_{2R} Q_{2L} F_1^\dagger
+f_{17}\overline{Q}_{2L} Q_{4R} S^\dagger
+f_{36}^*\overline{Q}_{4R} Q_{4L} F2
+ f_{7}\overline{Q}_{4L} D_{4R} H
+ M\overline{D}_{4R} D_{4L}     \nonumber\\
&&\hspace*{-20pt}
+ f_{50}\overline{D}_{4L} D_{5R} F_2^\dagger
+ f_{49}^*\overline{D}_{5R} D_{5L} F_3^\dagger
+ f_{26}\overline{D}_{5L} D_{2R} S
+ f_{48}^*\overline{D}_{2R} D_{2L} F_1
+ f_{15}\overline{D}_{2L} d_{2R} S
\label{twentytwod}
\ee
produces an effective coupling below the
TeV scale proportional to
\be
f_{10}f_{33}^*f_{17}f_{36}^*f_{7}f_{50}f_{49}^*f_{26}f_{48}^*f_{15}\frac{F_1^\dagger F_1 F_2^\dagger F_2 F_3^\dagger}{M^5}
\frac{(S^\dagger S)^2}{M^4}\overline{q}_{3L}d_{2R}H
+ h.c.\; .
\ee

Integrating out the heavy fermions in the tree level diagram
composed from the couplings
\be
&&\hspace*{-20pt}
f_{9}\overline{q}_{1L} Q_{1R} S^\dagger
+f_{32}^*\overline{Q}_{1R} Q_{1L} F_3
+f_{18}^*\overline{Q}_{1L} Q_{5R} S^\dagger
+f_{19}^*\overline{Q}_{5R} Q_{7L} S^\dagger
+ f_{8}\overline{Q}_{7L} D_{7R} H   \nonumber\\
&&\hspace*{-20pt}
+ M\overline{D}_{7R} D_{8L}
+ f_{30}\overline{D}_{8L} D_{6R} S
+ f_{28}\overline{D}_{6R} D_{1L} S
+ f_{14}\overline{D}_{1L} d_{1R} S
\label{twentytwoe}
\ee
produces an effective coupling below the
TeV scale proportional to
\be
f_{9}f_{32}^*f_{18}^*f_{19}^*f_{8}f_{30}f_{28}f_{14}\frac{F_3}{M}
\frac{(S^\dagger S)^3}{M^6}\overline{q}_{1L}d_{1R}H
+ h.c.\; .
\ee
There are four other very similar tree level contributions to $h_{11}^d$.

Integrating out the heavy fermions in the tree level diagram
composed from the couplings
\be
&&\hspace*{-20pt}
f_{9}\overline{q}_{1L} Q_{1R} S^\dagger
+f_{32}^*\overline{Q}_{1R} Q_{1L} F_3
+f_{18}^*\overline{Q}_{1L} Q_{5R} S^\dagger
+f_{19}^*\overline{Q}_{5R} Q_{7L} S^\dagger
+ f_{5}\overline{Q}_{7L} U_{6R} \bar{H}
+ f_{46}^*\overline{U}_{6R} U_{6L} F_1   \nonumber\\
&&\hspace*{-20pt}
+ f_{23}\overline{U}_{6L} U_{5R} S
+ f_{45}^*\overline{U}_{5R} U_{5L} F_2
+ f_{22}\overline{U}_{5L} U_{1R} S
+ f_{41}^*\overline{U}_{1R} U_{1L} F_3^\dagger
+ f_{12}\overline{U}_{1L} u_{1R} S
\label{twentytwof}
\ee
produces an effective coupling below the
TeV scale proportional to
\be
f_{9}f_{32}^*f_{18}^*f_{19}^*f_{5}f_{46}^*f_{23}f_{45}^*f_{22}f_{41}^*f_{12}\frac{F_1 F_2 F_3^\dagger F_3}{M^4}
\frac{(S^\dagger S)^3}{M^6}\overline{q}_{1L}u_{1R} \bar{H}
+ h.c.\; .
\ee

Integrating out the heavy fermions in the tree level diagram
composed from the couplings
\be
&&\hspace*{-20pt}
f_{9}\overline{q}_{1L} Q_{1R} S^\dagger
+f_{32}^*\overline{Q}_{1R} Q_{1L} F_3
+f_{18}^*\overline{Q}_{1L} Q_{5R} S^\dagger
+f_{19}^*\overline{Q}_{5R} Q_{7L} S^\dagger
+ f_{8}\overline{Q}_{7L} D_{7R} H
+ f_{52}^*\overline{D}_{7R} D_{7L} F_2 \nonumber\\
&&\hspace*{-20pt}
+ f_{29}\overline{D}_{7L} D_{5R} S
+ f_{49}^*\overline{D}_{5R} D_{5L} F_3^\dagger
+ f_{26}\overline{D}_{5L} D_{2R} S
+ f_{48}^*\overline{D}_{2R} D_{2L} F_1
+ f_{15}\overline{D}_{2L} d_{2R} S
\label{twentytwog}
\ee
produces an effective coupling below the
TeV scale proportional to
\be
f_{9}f_{32}^*f_{18}^*f_{19}^*f_{8}f_{52}^*f_{29}f_{49}^*f_{26}f_{48}^*f_{15}\frac{F_1 F_2 F_3^\dagger F_3}{M^4}
\frac{(S^\dagger S)^3}{M^6}\overline{q}_{1L}d_{2R}H
+ h.c.\; .
\ee

Integrating out the heavy fermions in the tree level diagram
composed from the couplings
\be
&&\hspace*{-20pt}
f_{9}\overline{q}_{1L} Q_{1R} S^\dagger
+f_{32}^*\overline{Q}_{1R} Q_{1L} F_3
+f_{18}^*\overline{Q}_{1L} Q_{5R} S^\dagger
+f_{19}^*\overline{Q}_{5R} Q_{7L} S^\dagger
+ f_{8}\overline{Q}_{7L} D_{7R} H
+ f_{52}^*\overline{D}_{7R} D_{7L} F_2 \nonumber\\
&&\hspace*{-20pt}
+ f_{29}\overline{D}_{7L} D_{5R} S
+ f_{50}^*\overline{D}_{5R} D_{4L} F_2
+ f_{24}^*\overline{D}_{4L} D_{3R} S
+ M\overline{D}_{3R} D_{3L}
+ f_3\overline{D}_{3L} d_{3R} S
\label{twentytwoh}
\ee
produces an effective coupling below the
TeV scale proportional to
\be
f_{9}f_{32}^*f_{18}^*f_{19}^*f_{8}f_{52}^*f_{29}f_{50}^*f_{24}^*f_{3}\frac{(F_2)^2 F_3}{M^3}
\frac{(S^\dagger S)^3}{M^6}\overline{q}_{1L}d_{3R}H
+ h.c.\; .
\ee
There is one other very similar tree level contribution to $h_{13}^d$.

Integrating out the heavy fermions in the tree level diagram
composed from the couplings
\be
&&\hspace*{-20pt}
  f_{11}\overline{q}_{3L} Q_{3R} S^\dagger
+ f_{34}^*\overline{Q}_{3R} Q_{3L} F_1^\dagger
+ f_{10}\overline{Q}_{3L} D_{3R} H
+ f_{24}^*\overline{D}_{3R} D_{4L} S^\dagger
+ f_{50}\overline{D}_{4L} D_{5R} F_2^\dagger
+ f_{29}^*\overline{D}_{5R} D_{7L} S^\dagger  \nonumber\\
&&\hspace*{-20pt}
+ f_{52}^*\overline{D}_{7L} D_{7R} F_2^\dagger
+ M\overline{D}_{7R} D_{8L}
+ f_{30}\overline{D}_{8L} D_{6R} S
+ f_{28}\overline{D}_{6R} D_{1L} S
+ f_{14}\overline{D}_{1L} d_{1R} S
\label{twentytwoi}
\ee
produces an effective coupling below the
TeV scale proportional to
\be
f_{11}f_{34}^*f_{10}f_{24}^*f_{50}f_{29}^*f_{52}^*f_{30}f_{28}f_{14}\frac{F_1^\dagger (F_2^\dagger )^2}{M^3}
\frac{(S^\dagger S)^3}{M^6}\overline{q}_{3L}d_{1R}H
+ h.c.\; .
\ee

The following effective couplings are not generated or are generated at
higher order in $\epsilon$ and/or $\beta$:
$h_{23}^u$, $h_{32}^u$, $h_{12}^u$, $h_{21}^u$, $h_{13}^u$, $h_{31}^u$,
and $h_{21}^d$. As already indicated these couplings are not needed
to reproduce the observed SM quark masses and mixings. For illustration,
$h_{32}^u$ arises from the effective coupling
\be
f_{11}f_{34}^*f_6f_{25}f_7^*f_{36}f_{17}^*f_2f_{42}^*f_{13}\frac{F_1^\dagger F_2 F_2^\dagger}{M^3}\frac{(S^\dagger S)^2}{M^4}
\frac{H^\dagger H}{M^2}\overline{q}_{3L}u_{2R}\bar{H} + h.c. \;,
\ee
so the extra suppression relative to eq. (\ref{ONE}) is by an additional factor
of $\beta$ as well as an additional factor of $\epsilon$.

Since $h_{12}^u$ and $h_{21}^u$ have extra suppression in this model,
$D^0 - \overline{D^0}$ mixing also has extra suppression. This weakens
the lower bound on $m_s$ derived in Section \ref{sec:3.2}.
Similarly since $h_{23}^u$ and $h_{32}^u$ have extra suppression the
relatively large BR for $t\rightarrow{c s}$ discussed in Section \ref{sec:4.2}
will not occur for this particular realization.

\section{Conclusion}

We have presented a framework in which only the top quark obtains its
mass from the Yukawa interaction with the SM Higgs boson via
dimension four operators.  All the other quarks receive their masses
from operators of dimension six or higher involving a complex scalar
$S$ that is part of an extended Higgs sector and 
whose vev is at the electroweak scale.  The successive hierarchy of
light quark masses is generated via the expansion parameter
$\left(\frac{S^\dagger S}{M^2}\right)\sim \epsilon^2$, where
$\epsilon \equiv \frac{v_s}{M} \sim 0.15$.  All the couplings of the
higher dimensional operators are of order one.  We are able to
generate the appropriate hierarchy of fermion masses with this small
parameter $\epsilon$.  Since $v_s$ is at the EW scale, the physics
of the new scale $M$ is not far above a TeV.  
We predict a neutral scalar $s$,
which gives rise to signals that could be detected
at the LHC or at the Tevatron. We make new predictions for
Higgs decays and for top decays. The model has a light
$Z'$ that has very weak couplings to SM fermions, but could be light enough
to be produced via mixing in Higgs decays at the LHC; this could give rise to
invisible Higgs decays, displaced vertices from the $Z'$
decays, or  multilepton final states, depending on the mass and lifetime of the $Z'$.

We have presented an explicit model in which the effective interaction given in (\ref{ONE})
is realized. This involves extending the SM gauge symmetry
by an abelian gauge symmetry $U(1)_S$ and a local flavon symmetry group
$U(1)_{F_1} \times U(1)_{F_2} \times U(1)_{F_3}$
The flavon symmetry is spontaneously broken at the TeV scale by a complex flavon scalars
$F_1$, $F_2$, $F_3$,
whereas the $U(1)_S$ symmetry is broken at the electroweak scale by the complex scalar $S$,
which is a SM singlet extension of the SM Higgs sector.
$S$ acts as the messenger of both flavor and electroweak symmetry breaking.
The model requires the existence  of vectorlike quarks and leptons,
both EW doublets and singlets, at the TeV scale.  These can be probed
at the LHC.  Their decays will be a new source for Higgs production and give rise to final states with
four $Z$'s or four $Z^\prime$'s and other interesting new physics signals at
the LHC.

We have restricted ourselves to models where all of the hierarchies of the
SM quark and charged lepton masses and mixings arise from powers of the
vev of a single messenger field. In \cite{bn}, a framework was suggested
in which all of these hierarchies arise from powers of
$\beta = \left(\frac{H^\dagger H}{M^2}\right)$. As we saw in the previous
section, in explicit models it is natural to generate powers of both
$\epsilon$ and $\beta$. Thus the model presented here and the framework
of \cite{bn} are two extremes of a more general class of models.
Obviously one could also generalize by introducing a more complicated
messenger sector, i.e. further extending the Higgs sector.

A truely viable model should have fewer species of heavy fermions than
were required in our example, ameliorating what is otherwise a
dramatic worsening of the little hierarchy problem of the Standard Model.
This could be achieved by a more efficient construction of the
messenger sector and its interplay with the flavon sector.
Another interesting direction is to attempt to generate some of
the higher order effective couplings from the top quark Yukawa, as
was done successfully with leptoquark-generated loop diagrams in \cite{Dobrescu:2008sz}.

\subsection*{Acknowledgements}

We are grateful to Bogdan Dobrescu for several useful discussions. SN
and ZM would like to thank the Fermilab Theoretical Physics Department
for warm hospitality and support during the completion of this work.
This research was supported in part by grant numbers
DOE-FG02-04ER41306 and DOE-FG02-ER46140.
Fermilab is operated by the Fermi Research Alliance LLC under contract
DE-AC02-07CH11359 with the U.S. Dept. of Energy.


\end{document}